# Uptake of stratospheric species on minerals proposed for stratospheric aerosol injection

Anais Lostier,[1] Yair Segev,[2,*] Tzemah Kislev,[2] Gal Schwartz Roitman,[2] Nadine Locoge,[1] Manolis N. Romanias.[1,*]

1: IMT Nord Europe, Institut Mines-Télécom, Univ. Lille, Centre for Energy and Environment, F-59000 Lille, France,

2: Stardust Labs Ltd., Ness Tziona, Israel 7403638

## Abstract

Solid mineral-based particles have been proposed as alternatives to sulfates for climate intervention by stratospheric aerosol injection, as a possible means for improving optical or chemical characteristics to thereby minimize risks and uncertainties. However, the heterogeneous reactivity of solid particles toward stratospheric trace gases, and possible implications to the ozone layer, is currently not fully constrained, particularly at stratospheric concentrations. Here we present a systematic comparative study of the uptake of nitric acid ($HNO_3$), hydrogen chloride (HCl), and nitrogen dioxide ($NO_2$) on four mineral surrogates, calcite, alumina, crystalline silica (quartz), and amorphous silica, using complementary Knudsen cell and flow-through reactor techniques. We find that $NO_2$ uptake is relatively weak on all surfaces, with estimated heterogeneous removal timescales indicating negligible direct impact on stratospheric nitrogen chemistry. Conversely, measuring HCl uptake over a concentration range spanning five orders of magnitude, we find substantial uptake with a pronounced concentration dependence consistent with surface site limited Langmuir adsorption. Extracting adsorption isotherms, we find that the surface coverage of HCl at stratospheric concentrations differs by four orders of magnitude between the surfaces, with calcite adsorbing the most and amorphous silica the least, suggesting a dominant role of surface acid-base character. Using HCl surface coverage as a proxy for the reactive uptake coefficient of $ClONO_2$, we estimate that amorphous silica could produce substantially lower ozone depletion due to chlorine activation than calcite or alumina under equivalent injection scenarios. We also find a marked difference in uptake between the crystalline and amorphous forms of silica, underscoring the sensitivity of heterogeneous chemistry to surface microstructure and the importance of selecting particles with low-reactivity surfaces, in addition to the consideration of bulk characteristics. Our findings motivate the development of particles with surfaces tailored for minimizing SAI risks and uncertainties, including minimal reactivity with stratospheric gases and background sulfate aerosols.

# 1. Introduction

Stratospheric aerosol injection (SAI) has emerged as a potential climate intervention strategy to temporarily offset anthropogenic warming by scattering a fraction of the incoming solar radiation back to space. While the capacity for global-scale cooling by SAI can be deduced from observations following large volcanic eruptions (e.g., Mt Pinatubo in 1991)[1], any SAI technology will also have to satisfy multiple stringent criteria to be sufficiently safe for sustained, large-scale deployment, including robust assessments of all foreseeable risks to the biosphere, methods for mitigating risks from uncertainties in climate side-effects such as stratospheric heating, cloud and precipitation patterns, and bounding of possible changes to the stratospheric composition, and to the ozone layer in particular[2,3].

The central element affecting the plausibility of meeting these requirements is the SAI particle. The option of increasing the amount of stratospheric sulfate aerosols[4–7], mimicking the cooling effects of large volcanic eruptions, has been widely studied in various modeling efforts[8–10] and preliminary practical assessments[11–14]. This option contains various uncertainties due to the complex time-evolution of a sulfate particle's size and composition as it transits the stratosphere[15], which inherently lead to particle size distributions that depend on the total aerosol burden, its latitudinal distribution and geographic deployment strategy. These in turn make it difficult to bound even some known side-effects of sulfates – namely, stratospheric heating due to infrared absorption, which could disrupt atmospheric circulation patterns[16,17], global ozone depletion due to conversion of reactive nitrogen oxides that participate in chlorine deactivation[18–20], and enhancement of polar ozone loss[21], as these effects depend on the sulfate particle volume and surface area. An additional limitation of sulfates as a building block for SAI may be that small-scale testing, which will be necessary before any large-scale deployment is considered, is impractical due to the high natural background levels already in the lower stratosphere (hundreds of thousands of tons in volcanically quiescent periods)[22]; this is especially limiting if deployment is based on slow processes such as outdoor conversion of precursor sulfur dioxide, which would require a large amount of material for any observability.

These drawbacks have motivated the search for alternative particles, especially ones based on solid mineral dust such as aluminum oxide ($Al_2O_3$, alumina)[23–28], calcium carbonate ($CaCO_3$, including calcite or other polymorphs)[27–32], and silicon dioxide ($SiO_2$, including quartz or other forms of silica)[23,25,30,33,34]. These minerals are generally abundant in the troposphere but practically absent in the stratosphere, providing a relatively good baseline of knowledge on their ecological impacts without a background-level limitation on stratospheric observability. The use of alternative materials as building blocks for SAI particles also opens the door to optimizing optical properties and harnessing existing manufacturing capabilities[35]; however, substantial uncertainties remain regarding the atmospheric behavior of such materials. In particular, while the bulk material dominates the optical properties of particles, reactivity is controlled by the surface, which can participate in heterogeneous reactions that could impact the stratospheric composition.

Relatively non-reactive surfaces including some minerals can potentially promote stratospheric ozone depletion through chlorine activation by facilitating the heterogeneous reaction between the reservoirs HCl and $ClONO_2$, the main reaction responsible for chlorine activation on particles composing polar stratospheric clouds (PSCs)[36–38]. Adsorption of HCl initiates the reaction, and the specific adsorption mechanism will partially determine the reactivity; for example, dissociative adsorption of HCl, as opposed to reversible physical adsorption, creates a chlorine radical with a large reactive cross-section[26,27,39–41]. Calcite may also react directly with HCl to form $CaCl_2$, a mechanism which has been suggested as a possible route for chlorine removal from the stratosphere[29], though surface passivation over time might reduce the effectiveness[31]. Competition between HCl and other gas species, such as nitric acid ($HNO_3$), may limit the availability of adsorption sites and reduce the chlorine activation rate[42]. Nitric acid uptake, however, may also enhance ozone depletion if it is removed from the stratosphere on the surface of settling particles, as it is a reservoir for reactive nitrogen dioxide ($NO_2$), which participates in chlorine deactivation; this denitrification mechanism contributes to the impact of PSCs[43–45]. In addition, the presence of water vapor, even at the very low concentrations characteristic of the stratosphere, can influence the uptake of acidic gases on particle surfaces, particularly under low-temperature conditions. Water may either inhibit or promote uptake by competing for adsorption sites or by facilitating dissociative adsorption pathways, thereby modifying heterogeneous reaction rates[46]. Taken together, the contributions of various physical, reactive, and competitive co-adsorption mechanisms are strongly dependent on gas composition and concentrations, requiring appropriate experimental investigation to accurately inform any modeling studies on SAI impacts.

Laboratory investigations of kinetic and partitioning parameters can quantify uptake coefficients, adsorption capacities, and gas-surface interaction mechanisms between stratospherically relevant trace gases and particle surfaces, providing key inputs for model-based assessments of particle aging, heterogeneous loss processes, and impacts on stratospheric composition. Common reactor systems for such measurements include coated-wall flow tubes, aerosol flow tubes, and related configurations. Coated-wall flow tubes are suitable for investigation of kinetics over long timescales under atmospheric conditions; however, these systems can be affected by gas-phase diffusion limitations when surface uptake is efficient, while uncertainties in accessible surface area and coating morphology may influence quantitative interpretation. Aerosol flow tubes better simulate airborne particles, but their relatively short gas–particle interaction times - typically ranging from seconds to a few minutes - can limit their ability to investigate long-term particle aging processes, which are particularly relevant for SAI applications given the expected stratospheric residence times of particles. Their relatively low surface-to-gas ratios and downstream sampling requirements may introduce additional uncertainties, particularly under low-temperature conditions where adsorption and desorption processes can be strongly temperature dependent. In this context, Knudsen reactors remain particularly well suited for investigating stratospherically relevant heterogeneous chemistry[47]. Initially applied in atmospheric chemistry to examine reactions on ice surfaces relevant to polar stratospheric cloud chemistry, these systems operate under molecular flow conditions[38], eliminating gas-phase diffusion limitations and enabling direct investigation of gas-surface interactions driven solely by the intrinsic affinity of trace gases for the substrate. Their ability to operate under stratospherically relevant temperatures makes them particularly suitable for probing the long-term reactivity of candidate SAI materials. Importantly, kinetic data derived from Knudsen reactor experiments have been widely incorporated into major atmospheric kinetic evaluations, including those conducted by the NASA Jet Propulsion Laboratory (JPL) and IUPAC, highlighting their recognized relevance for atmospheric chemistry modeling[48,49].

In recent years, increasing attention has been devoted to the heterogeneous reactivity of inorganic trace gases on candidate SAI materials using various experimental approaches, particularly flow tube reactors, often coupled with *in situ* spectroscopic techniques to elucidate reaction pathways and surface transformations[31,34,50]. There has also been significant recent progress in implementing mechanistic insights of heterogeneous reactions into SAI impact assessments[27,28]. However, high fidelity experimental data required for modeling the chemical impacts of proposed minerals specifically under stratospheric concentrations, which can significantly affect uptake processes, remains limited. Furthermore, no study has yet presented systematic comparative measurements across multiple material candidates using consistent experimental approaches.

Here we present a comprehensive experimental investigation of heterogeneous uptake and reactivity for four solid mineral surrogates of current interest for stratospheric aerosol injection applications: calcite, alumina, crystalline silica (quartz), and amorphous silica. Using complementary Knudsen and flow-through reactor (packed bed configuration) techniques, we measure uptake coefficients for three stratospherically relevant trace gases, nitric acid, hydrogen chloride, and nitrogen dioxide, across wide concentration ranges to enable reliable extrapolation to atmospheric conditions. Our measurements reveal significant concentration dependencies in uptake behavior, provide detailed adsorption isotherms for surface capacity assessment, and identify important differences in reactivity patterns between the surfaces that directly inform the evaluation of these materials as potential SAI candidates.

# 2. Materials and methods

## 2.1 Mineral surfaces

The solid samples used in this study, calcite ($CaCO_3$), alumina ($Al_2O_3$), quartz (crystalline silica, $SiO_2$), and amorphous silica, were selected based on their relevance as proposed building blocks for potential SAI particles. This selection covers a range of chemical compositions and surface properties that influence reactivity with stratospheric trace gases. All samples were used as received and stored under dry conditions. Details of the samples, including specific surface area (*SSA*) as defined by the Brunauer-Emmett-Teller method (BET), are provided in **Table 1**.

Calcium carbonate has been extensively studied for its reactions with stratospheric acids such as $HNO_3$ and HCl, leading to the formation of $Ca(NO_3)_2$ and $CaCl_2$, both of which can further participate in ozone-depleting reactions under stratospheric conditions[31,51]. Aluminum oxide has also emerged as a viable SAI candidate due to its favorable radiative properties and relatively low heterogeneous reactivity, although recent assessments have highlighted potential uncertainties in its chemical impact on the stratospheric ozone layer[24,26]. Quartz, though less favored in models of SAI scenarios due to its infrared absorption in the atmospheric longwave window, has been included here to compare reactivity across different mineral classes. Its heterogeneous chemistry with nitrogen and chlorine oxides has been well documented[34,52]. Amorphous silica, though ubiquitous in industry and in several natural forms, has not been discussed as a candidate SAI particle. The surface properties of amorphous silica may differ substantially from crystalline silica, leading to differences in uptake behavior. In this study we included a specific surrogate of amorphous silica, fumed silica powder.

**Table 1:** List of samples selected with information on the measured BET SSA, supplier and purity.

| **Name** | **Formula** | **Purity (%)** | **BET *SSA* ($m^2$ $g^{-1}$)** | **Supplier** |
|---|---|---|---|---|
| Crystalline silica powder (<63 µm) | $SiO_2$ quartz | > 98.7 | 1.5 ± 0.1* | Sigma-Aldrich (83340) |
| Amorphous silica powder (14 nm‡) | $SiO_2$ fumed | > 99.8 | 200 ± 25† | Evonik (Aerosil® 200) |
| Calcite nanoparticles (90±15 nm) | $CaCO_3$ | > 98 | 21.1 ± 0.5* | PlasmaChem (PL-CACOU) |
| Alumina powder (1 µm) | α-$Al_2O_3$ | n.a. | 1.6 ± 0.2* | Schmitz Metallographie (SKU: 150-300) |

*: Measured; †: provided by the supplier; n.a: not available; ‡ primary particle size.

## 2.2 Gas synthesis

Nitric acid ($HNO_3$) was generated from a liquid-phase mixture of concentrated $H_2SO_4$ and $HNO_3$ maintained at 281 K to reduce the vapor pressure of $HNO_3$ and control its release. Gas purity exceeded 99% as confirmed by quadrupole mass spectrometry monitoring of the $m/z$ = 46 and $m/z$ = 30 ratio (1.8), consistent with literature values[53]. $HNO_3$ concentrations used in this study ranged from $2\times10^{12}$ to $3\times10^{13}$ molecules $cm^{-3}$.

Hydrogen chloride (HCl) was synthesized by reacting KCl with concentrated $H_2SO_4$, followed by condensation at 77 K and vacuum purification. Mass spectrometric analysis confirmed >99% purity and appropriate isotope ratios. Neat HCl (4 L glass bulb) supplied concentrations of $2\times10^{11}$ to $9\times10^{13}$ molecules $cm^{-3}$, with diluted mixtures (4.8% in He) providing lower concentrations down to ~$7\times10^{9}$ molecules $cm^{-3}$, compared to typical values of ~$2\times10^{9}$ molecules $cm^{-3}$ at an SAI relevant altitude of 20km[54].

Nitrogen dioxide ($NO_2$) was introduced from a glass bulb filled from a commercial cylinder (>99% purity). In the U-shaped reactor, the gas was diluted in helium, and the concentration was further diluted and controlled by mixing with zero-air using mass flow controllers. $NO_2$ concentrations used in this study ranged from $8\times10^{11}$ to $1\times10^{14}$ molecules $cm^{-3}$.

Gas preparation and synthesis protocols are detailed in Supporting Information section **S.1**.

## 2.3 Experimental setups

We used two complementary experimental systems to investigate heterogeneous interactions between the selected trace gases and mineral surfaces: a low-pressure Knudsen cell reactor and an atmospheric-pressure U-shaped flow-through reactor. The Knudsen cell reactor operates under molecular flow conditions, where gas-phase diffusion limitations are negligible.

This configuration is used to investigate heterogeneous kinetics and enables accurate determination of kinetic parameters on millisecond timescales, such as initial uptake coefficients, while also allowing prolonged surface exposure for several minutes to hours to derive steady-state uptake coefficients and adsorption isotherms. In parallel, a U-shaped reactor operating at atmospheric pressure under slow-flow conditions was used to probe comparatively slower processes, providing measurements of steady-state uptake coefficients and adsorption behavior. The Knudsen cell was used to determine uptake coefficients of HCl on all four substrates, as well as the uptake of $HNO_3$ on alumina and calcite, with $HNO_3$ serving as a reference species to assess the consistency of the setup with literature data. Uptake of $NO_2$ was investigated with both setups to cross-validate the measurements between the two experimental setups.

### 2.3.1 Knudsen cell measurements

The Knudsen cell operated at a controlled temperature ("room temperature", 296 K) under molecular flow conditions, where molecule-surface interactions are governed purely by gas-surface affinity without interference from carrier gas effects. This widely adopted approach, described by Caloz et al., provides high-quality kinetic data, which have been included in major IUPAC and NASA JPL evaluations[47–49].

The Pyrex glass reactor (1,830 $cm^3$ total volume, Teflon-coated) contained samples at the bottom with gas introduced from the top (see Supporting Information **Figure S1**). A movable plunger controlled the surface exposure, while molecules escaped through orifices of various selectable diameters (14, 8, 4, and 1 mm), enabling residence time control. Escaping molecules entered a differentially pumped mass spectrometer chamber where the entering beam was modulated at 140 Hz by a chopper and detected by a quadrupole mass spectrometer (QMS, Balzers QMG 421, 70 eV). The chopper frequency was synchronized with a lock-in amplifier (SR830 DSP), ensuring only molecules exiting the cell and forming the molecular beam were detected (phase-sensitive detection). The Lock-in signal was digitized and recorded for analysis.

This modulation approach is critical for sticky species like HCl and $HNO_3$, eliminating wall memory effects that can lead to significant underestimation of uptake coefficients. Further information and examples for the importance of signal modulation for the determination of accurate kinetic parameters of sticky species are provided in Supporting Information section **S.2.1.1** and **Figure S2** therein. Additional information about the Knudsen cell may be found elsewhere[55,56].

**Experimental protocol**: Samples are pumped down under vacuum for 20 min at 296 K before exposure, to remove adsorbed water and other volatile species. This pretreatment is carried out using the largest escape orifice (14 mm) to facilitate desorption. Once this step is completed, the plunger is moved down to isolate the surface, and the background signal of gas is recorded. The mass spectrometric signal of the gas is monitored at characteristic *m*/*z* values ($HNO_3$: 63, HCl: 36, $NO_2$: 46). A constant flow of gas is then introduced into the cell. Once steady-state conditions are established, indicated by a stable lock-in signal, the plunger is raised to expose the sample to the gas. After the uptake phase, the gas flow is stopped, and the signal returns to background levels. The plunger is then raised again to monitor any desorption of gas from the solid surface. This protocol was followed for all uptake coefficient measurements. However, in the case of surface coverage determination, cumulative uptakes were performed at different concentrations on the same substrate, and the desorption step was performed at the end of the process. All uptake measurements were performed under dry conditions with estimated background water concentration levels <$10^{10}$ molecules cm$^{-3}$.

**Uptake coefficient determination**: The initial uptake coefficient characterizes the interaction of a gas with a pristine surface and is determined within the first milliseconds of gas exposure[49]. It provides insight into the most reactive surface sites and helps correlate surface composition with gas uptake behavior. The Knudsen cell allows for short residence times and rapid detection. By maximizing the orifice size (i.e. 14 mm), the $HNO_3$, $NO_2$, and HCl residence time is reduced to 0.26, 0.22, and 0.20 s, respectively, enhancing sensitivity to initial uptake. The initial uptake coefficient, $\gamma_0$, is calculated using Equation 1:

$$\gamma_0 = (\frac{I_0}{I_r} - 1) \times \frac{4 \times V_r \times k_{esc}}{c \times A_s} \qquad \text{Eq.1}$$

where $I_0$ and $I_r$ are the pre-exposure and initial exposure signals respectively, $V_r$ is the cell volume, $k_{esc}$ is the escape rate of the molecules, related to their molecular weight and the geometric characteristics of the reactor, $c$ is the mean molecular velocity, and $A_s$ is the geometric surface area. We assume that during this brief interval only part of the dust surface is exposed, so $A_s$ is taken as the area of the sample holder on which the powder is placed; thus, the extracted value of $\gamma_0$ represents an upper limit.

Steady-state uptake coefficients ($\gamma_{ss}$) were determined after prolonged exposure, determined when the signal varied by less than 3% over a 15-minute period:

$$\gamma_{ss} = (\frac{I_0}{I_{ss}} - 1) \times \frac{4 \times V_r \times k_{esc}}{c \times m \times SSA} \qquad \text{Eq.2}$$

where $I_{ss}$ is the final exposure signal, $m$ is the sample mass, and $SSA$ is the specific surface area. Here we use the BET surface area, assuming gas molecules access the entire available surface under long-term exposure conditions. This assumption has been validated in previous studies and was further confirmed in the present work by varying the sample mass and independently determining $\gamma_{ss}$ values normalized to $SSA$, representing the highest available surface area of a material for uptake[49]. Therefore, the reported $\gamma_{ss}$ values referenced to $SSA$ provide a lower limit of the measured $\gamma_{ss}$.

**Adsorption isotherms determination:** For heterogeneous reactions involving one gas species, e.g. $ClONO_2$, and another adsorbed species, e.g. HCl, the rate limiting factor in the low-concentration regime is typically the surface coverage of adsorbed molecules, i.e. the number of molecules taken up per unit surface area. This parameter is obtained by integrating the net molecular flux of gas consumed by the surface during exposure, *F(t)* (see **Figure S3**). When a non-zero steady-state gas consumption $F_{ss}$ is observed, indicating that the surface maintains a sustained uptake capacity due either to reactive regeneration of sites or to very slow diffusion into the bulk solid, the integration of the adsorption peak is performed until the steady-state regime is reached, and excluding the steady state contribution (i.e. only the shaded region shown in **Figure S3**). The resulting value corresponds to the steady-state surface coverage, $N_{ss}$:

$$N_{ss} = \int_{t=0}^{t} \frac{F(t)-F_{SS}}{m \times SSA} \qquad \text{Eq 3.}$$

This approach is commonly used in the literature because it provides a measure of uptake capacity that is independent of the total exposure duration for reactive systems, and is conceptually consistent with the two-thirds criterion proposed by Abbatt et al[46,57].

Measured curves of the surface coverage as a function of the concentration were fitted using the Langmuir isotherm model to determine the equilibrium constant ($K_{Lang}$), the maximum steady state adsorption capacity ($N_{ss,max}$), and their product, the partitioning coefficient ($K_{Lin}$), which the IUPAC task group has recommended as a standardized comparison parameter for different mineral surrogates[49]. Finally, the amount of reversibly adsorbed molecules is measured by integrating the signal during the desorption phase of each experiment, when the flow is turned off, and normalizing to $N_{ss}$, providing the recovered fraction $\%N_{rev}$.

### 2.3.2 U-shaped reactor measurements

The U-shaped reactor enabled steady-state $NO_2$ uptake measurements under atmospheric pressure for cross-validating the Knudsen cell results. Dry zero air was used as the bath gas. The system comprised two 15 $cm^3$ sections connected by 3-way valves: an upper bypass line for concentration monitoring and a lower section containing fritted glass sample holders (see **Figure S4**). A selected-ion flow tube mass spectrometer (SIFT-MS, Voice 200) was employed for $NO_2$ detection (4 s time resolution) using $O_2^+$ reagent ions at $m/z$ = 46. Additional information about the setup is provided in Supporting Information section **S.2.2**.

Experiments involved initial bypass concentration measurement, flow direction through the sample-containing section until steady-state equilibrium, and return to bypass for concentration verification. The steady-state uptake coefficient was determined by:

$$\gamma_{SS} = \frac{4 \times (d\{NO_2\}/dt)}{c \times m \times SSA \times [NO_2]_0} \qquad \text{Eq 4.}$$

where $d\{NO_2\}/dt$ is the steady state uptake rate and $[NO_2]_0$ is the introduced concentration[58].

## 2.4 Uncertainty analysis

Uncertainties were determined using quadrature propagation of systematic errors including gas purity (1%), flow measurements (5%), signal precision (2-5% depending on concentration), and surface area determination (2.4-12.5%). Additional sources included escape rate measurements (1-10%, depending on orifice size) and signal integration uncertainties (10% for $N_{ss}$ values). Overall uncertainties for Knudsen cell measurements were: $\gamma_0$ ±13%, $\gamma_{ss}$ ±18%, and $N_{ss}$ ±20%. It should be noted that for $NO_2$ concentrations below $1 \times 10^{13}$ molecules $cm^{-3}$ the uncertainties in $N_{ss}$ were on the order of 50% owing to the weak uptake of $NO_2$ on the mineral surrogates. U-reactor measurements achieved $\gamma_{ss}$ uncertainties of ±18%. Reproducibility between independent experiments was consistently better than 5%. Additional details of the uncertainty analysis methods are provided in Supporting Information section **S.2.3**.

# 3. Results

## 3.1 $HNO_3$ uptake on mineral surrogates

Before focusing on the uptake of HCl and $NO_2$, Knudsen cell measurements of $HNO_3$ are conducted on the samples for benchmarking, as its uptake on mineral surrogates has been extensively studied in the literature using various experimental approaches, with results evaluated by the IUPAC panel[49]. For this gas species, our measurements targeted the initial uptake. Notably, the initial uptake coefficients of $HNO_3$ have been found in the literature to be independent of the mass of exposed mineral particles. Consequently, the IUPAC panel recommends expressing these values based on the geometric surface area of the material. **Figure S5** presents literature data for $HNO_3$ initial uptake on $Al_2O_3$ and $CaCO_3$, obtained using various Knudsen cell reactors and aerosol flow tubes. There is good agreement between the literature values in the concentration range studied except for two studies which should be excluded, as discussed in section **S.3.1** in the Supporting Information.

Several studies have reported a dependence of the initial uptake coefficient on $HNO_3$ concentration, with values decreasing as the concentration increases[53,59–64]. Hanisch and Crowley attribute this behavior to $HNO_3$ desorption from reactor walls, which counteracts its rapid uptake and leads to observably lower uptake coefficients[61]. Although the potential saturation of surface sites at higher concentrations has not been explicitly considered, it cannot be ruled out, especially if $HNO_3$ uptake is limited to the geometric surface of the particles, as this could reduce the reaction probability even during the initial uptake phase.

Our results, along with their associated uncertainties, show excellent agreement with previous studies, demonstrating the reliability of our experimental setup. Furthermore, the strong agreement between uptake coefficients obtained across different experimental setups highlights the robust consistency of these measurements. By varying the $HNO_3$ concentration by a factor of 12 in the concentration range of $10^{11}$ to $10^{12}$ molecules $cm^{-3}$, no significant concentration dependence is observed, indicating that the Knudsen cell reactor walls are inert and do not contribute measurably to the initial uptake coefficient. Overall, the initial uptake coefficients of $HNO_3$ on these mineral surrogates are high (on the order of 0.1), indicating strong partitioning into the adsorbed phase. Across the concentration range of $10^{10}$ to $10^{13}$ molecules $cm^{-3}$, encompassing both the present study and literature data, a slight concentration dependence is observed, and the data are described using an empirical expression (see Supporting Information section **S.3.1**).

## 3.2 HCl uptake on mineral surrogates

Hydrogen chloride was found to strongly interact with all four mineral surrogates. Unlike the literature, where mineral particles were exposed to HCl for only a few seconds to minutes[51,53], this study extended exposure times from several minutes to hours. The experiments were conducted over a wide concentration range, and the key parameters determined included the initial and steady-state uptake coefficients, adsorption isotherms, and the reversibly adsorbed fraction. Results are summarized in **Table 2**, and a representative uptake profile of HCl on alumina is shown in **Figure S3**.

The initial uptake coefficients of HCl, normalized to the geometric area, $\gamma_0(geom)$, were found to be independent of the initial concentration, and ranged between 0.01 and 0.43 for the different materials, in the following order: $SiO_2$ fumed < $SiO_2$ quartz < $Al_2O_3$ < $CaCO_3$. These $\gamma_0(geom)$ values are relatively high, indicating a strong affinity of HCl to the surface sites of the minerals.

**Table 2.** Summary of experimental results on HCl uptake on mineral surrogates, at room temperature.

| Material | $\gamma_0$(geom) | $\gamma_{ss}$(BET); [HCl] in molecules cm$^{-3}$ | $N_{ss,max}$ (molecules cm$^{-2}$) | $K_{lang}$ (cm$^{-3}$ molecule$^{-1}$) | $K_{lin}$ (cm) | % $N_{rev}$ |
|---|---|---|---|---|---|---|
| $SiO_2$ fumed | $(1.4\pm0.2)\times10^{-2}$ | $7.6\times10^{2}\times[HCl]^{-0.85}$ | $(1.5\pm0.4)\times10^{12}$ | $9.3\times10^{-14}$ | 0.14 | 37.4 |
| $SiO_2$ Quartz | $(5.2\pm0.7)\times10^{-2}$ | $6.1\times10^{7}\times[HCl]^{-1.16}$ | $(3.5\pm0.7)\times10^{13}$ | $1.4\times10^{-14}$ | 39.6 | 24.1 |
| $Al_2O_3$ | $0.38\pm0.05$ | $8.27\times10^{8}\times[HCl]^{-1.20}$ | $(1.9\pm0.4)\times10^{14}$ | $1.2\times10^{-12}$ | 228 | 32.8 |
| $CaCO_3$ | $0.43\pm0.06$ | $2.7\times10^{7}\times[HCl]^{-1.10}$ | $(3.0\pm0.6)\times10^{14}$ | $1.6\times10^{-12}$ | 480 | 7.8 |

As shown in **Figure 1**, the steady-state uptake coefficient normalized to the BET specific surface area, $\gamma_{ss}(BET)$, increases as the HCl concentration decreases, rendering the uptake process non-first order. This trend, observed by varying the concentration over four orders of magnitude, can be attributed to the limited number of available surface sites, and similar observations have been reported in the literature[55,65].

Empirical expressions for the measured uptake coefficient, presented in **Table 2**, reproduce the experimental data over five orders of magnitude in concentration, enabling extrapolation to stratospherically relevant concentrations. The observed concentration dependence and non-first-order behavior of the steady-state uptake coefficients suggests that the interaction of HCl with mineral surfaces may involve a more complex mechanism, potentially comprising multiple or consecutive processes. Among all materials tested, amorphous silica exhibited the lowest $\gamma_{ss}(BET)$ values for atmospherically relevant concentrations, while alumina and calcite showed the highest. These trends likely arise from differences in chemical composition and uptake mechanisms. For instance, HCl uptake on calcite is reactive (see below), whereas on amorphous silica it is probably non-reactive or less reactive. HCl is an acidic gas and is therefore expected to interact more readily with alkaline surfaces such as $CaCO_3$ through acid–base interactions. In contrast, amorphous silica generally exhibits an acidic surface[66], making strong interactions with an acidic gas less favorable. This difference in surface chemistry further explains the enhanced uptake observed for $CaCO_3$. An extended discussion of these measurements and their comparison to literature is presented in Supporting Information section **S.3.2.1**.

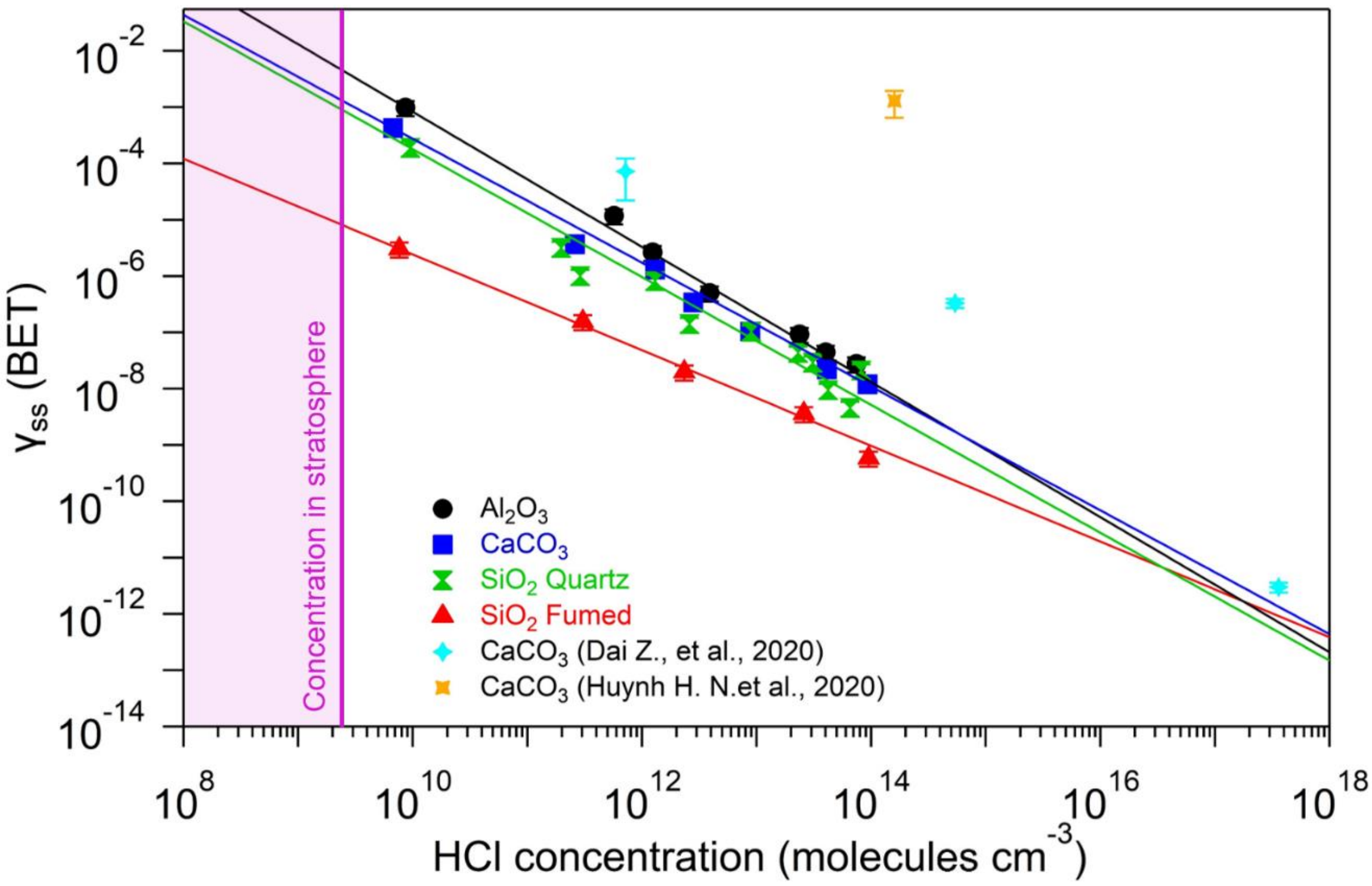

**Figure 1.** Steady-state uptake coefficients, $\gamma_{ss}(BET)$, of HCl on various mineral surfaces as a function of HCl gas-phase concentration, determined at room temperature in this work and low temperatures in literature studies[31,51]. Our experimental data are fitted using empirical expressions across a broad range of concentrations. For clarity, the error bars displayed for results from the current study correspond to ±30% uncertainty.

The adsorption isotherms for HCl, obtained over a wide concentration range, are presented in **Figure 2**. In all cases, the uptake behavior follows a Langmuir-Hinshelwood reactive model, where $N_{ss}$ reaches a saturation value as HCl concentration increases. Fitting the experimental data to a Langmuir model, which assumes equivalent surface sites, can be justified for the current system since HCl, as an acid, likely exhibits a selective affinity for sites with basic characteristics (**Table 2**).

Among the studied materials, amorphous silica exhibited the lowest uptake capacity with saturation levels, $N_{ss,max}$, around $10^{12}$ molecules cm$^{-2}$, followed by quartz, with $N_{ss,max}$ values an order of magnitude higher. Alumina and calcite showed the highest uptake capacities, reaching saturation levels on the order of $10^{14}$ molecules cm$^{-2}$. The HCl partitioning coefficient for calcite was found to be >$10^4$ times greater than that of amorphous silica, more than 30 times greater than that of quartz, and approximately a factor of two greater than that of alumina (**Table 2**).

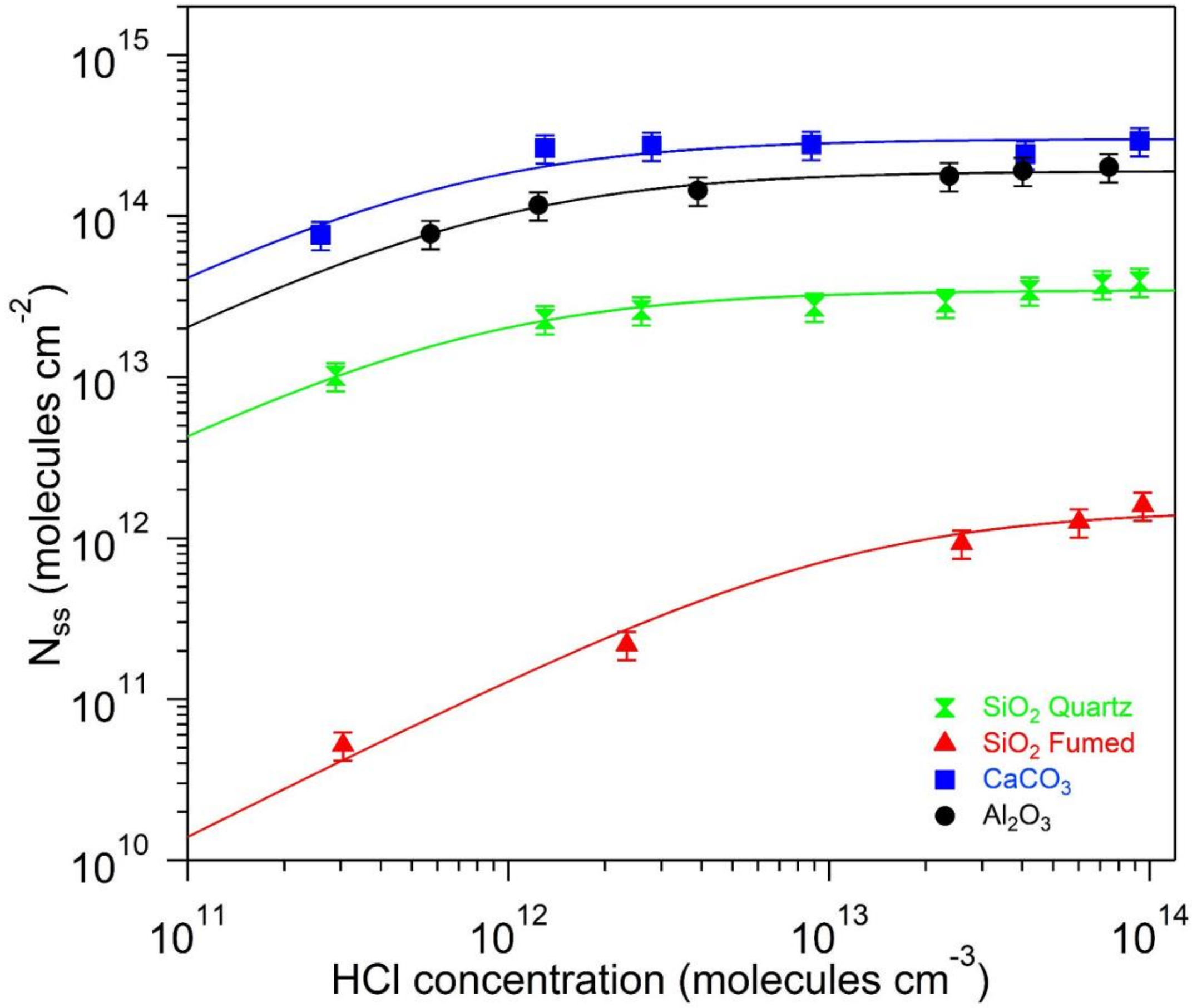

**Figure 2**. Adsorption isotherms of HCl on various mineral surrogates, determined at room temperature. Experimental results are fitted using the Langmuir adsorption isotherm model. Error bars correspond to $N_{ss}$ uncertainty of ±20% as given in section **2.4**.

In all cases, part of the HCl is reversibly adsorbed, as was confirmed by observing its desorption. Besides calcite, the desorbed fraction $N_{rev}$ ranged between 24% and 37%. The remaining HCl fraction that was taken up can either be strongly bound on the surface (unreactive) or have reacted, leading to surface or gas-phase products. Given that the temperature dependence of desorption on temperature is a negative exponential, HCl desorption from mineral surrogates is expected to be minimal under stratospherically relevant temperatures.

In the Knudsen cell, identifying surface products is particularly challenging, as detection is only possible if these species desorb. However, gas-phase reaction products can be directly detected using mass spectrometry. The reaction product expected from the reaction of HCl with alumina or silica is water, for example:

$$Al_2O_3 + 6HCl \rightarrow 2AlCl_3 + 3H_2O \qquad \text{(R1)}$$

For both forms of silica, no water traces were detected in the gas phase, which could indicate that the uptake is non-reactive. In contrast, traces of water were detected when alumina was exposed to HCl, confirming that part of the uptake is reactive. However, since water has a strong affinity for the surface, its quantification remains challenging, making it difficult to determine the absolute reaction yield, and we cannot exclude that HCl uptake on silica is partially reactive.

The uptake of HCl on $CaCO_3$ includes a known reactive component, resulting in the formation of $CO_2$ and $H_2O$, with a simplified representation:

$$CaCO_3 + 2HCl \rightarrow CaCl_2 + CO_2 + H_2O \qquad (R2)$$

According to Santschi and Rossi[53], the $CaCO_3$ interface is mainly structured as a bifunctional intermediate $Ca(OH)(HCO_3)$, and the reaction proceeds through several steps, first involving the physical adsorption of HCl on the surface and then its reaction with the OH surface groups to finally yield $CO_2$ and $H_2O$, in the following sequence of reactions:

$$Ca(OH)(HCO_3) + HCl_{(ads)} \rightarrow CaCl(HCO_3) + H_2O \qquad (R2a)$$

$$CaCl(HCO_3) + HCl_{(ads)} \rightarrow CaCl_2 + CO_2 + H_2O \qquad (R2b)$$

This reaction mechanism follows a Langmuir-Hinshelwood behavior, in accordance with our experimental observations. Given that $CO_2$ is highly volatile, it readily desorbs into the gas phase, allowing it to be monitored in parallel with the HCl concentration, as presented in **Figure S6**. The initial rapid decrease in HCl concentration upon exposure to the surface is followed by a delayed release of $CO_2$ into the gas phase. This delay suggests that $CO_2$ formation is directly linked to HCl surface coverage, which is consistent with the reaction stoichiometry, as two HCl molecules are required for the formation of one $CO_2$ molecule. A critical surface coverage threshold appears to exist, above which $CO_2$ formation becomes linearly dependent on HCl consumption (**Figure S7**).

The total consumed HCl molecules, determined by the ratio of the integrated HCl uptake peak and that of $CO_2$ formation, is observed to decrease with increasing HCl gas concentration, see Supporting Information section **S.3.2.2** and **Table S1**. According to our observations, under stratospherically relevant HCl concentrations, the reactive fraction of HCl molecules taken up by the calcite surface can fall below 20%, indicating that a substantial portion of the adsorbed HCl remains unreacted and may be available to interact with other species. These observations highlight the critical role of surface coverage in governing HCl reactivity.

## 3.3 $NO_2$ uptake on mineral surrogates

The uptake of nitrogen dioxide on mineral surrogate surfaces was investigated using both the Knudsen cell and the U-shaped reactor. The objective was to cross validate the measurements, which were conducted over a broad concentration range ($7\times10^{11}$ to $6\times10^{13}$ molecules $cm^{-3}$) at room temperature, and the results are summarized in **Table 3**. It should be noted that $NO_2$ uptake was found to be irreversible for all surfaces, pointing to a reactive process or strong bonding with the corresponding surface site.

The initial uptake coefficient $\gamma_0(geom)$, measured using the Knudsen cell, was within the same order of magnitude for all materials, ranging from $5\times10^{-3}$ to $10^{-2}$. These values are significantly lower than those reported for HCl and $HNO_3$, suggesting a weaker interaction of $NO_2$ with the mineral surfaces. Considering that the initial uptake typically reflects a selective interaction with specific surface sites, it is likely that $NO_2$ targets a limited number of reactive sites common to all mineral substrates.

**Table 3.** Summary of experimental results for $NO_2$ uptake on mineral surrogates at room temperature. For surfaces where a concentration dependence of $\gamma_{ss}(BET)$ is denoted, the expressions were obtained by combining measurements from the Knudsen cell and the U-shaped reactor. The steady-state uptake coefficients for amorphous silica and calcite were determined at concentrations on the order of 5 × $10^{13}$ molecules $cm^{-3}$ and no concentration dependence was investigated.

| **Material** | **$\gamma_0$ (Geom)** | **$\gamma_{ss}$ (BET)** | **$N_{ss,max}$ (molecules $cm^{-2}$)** | **$K_{lang}$ ($cm^3$ $molecule^{-1}$)** | **$K_{lin}$ (cm)** |
|---|---|---|---|---|---|
| $SiO_2$ fumed | $(8.7±2.6)×10^{-3}$ | $1.6 × 10^{-9}$ | $5.2×10^{11}$ | $4.2×10^{-14}$ | 0.022 |
| $SiO_2$ Quartz | $(5.1±0.7)×10^{-3}$ | $4.0×10^{-9} + 1.73×10^7[NO_2]^{-1.15}$ | $3.3×10^{12}$ | $4.2×10^{-13}$ | 1.37 |
| $Al_2O_3$ | $(1.3±0.4)×10^{-2}$ | | $3.1×10^{12}$ | $5.0×10^{-14}$ | 0.16 |
| $CaCO_3$ | $(6.3±3.2)×10^{-3}$ | $1.8 × 10^{-8}$ | $2×10^{12}$ | $2.0×10^{-14}$ | 0.04 |

The steady-state uptake coefficients $\gamma_{ss}(BET)$ of $NO_2$ on various mineral surfaces have been widely reported in the literature and summarized by the IUPAC Task Group in their most recent evaluation[49]. Most measurements were conducted at $NO_2$ concentrations of $10^{13}$ molecules $cm^{-3}$ or higher, yielding $\gamma_{ss}(BET)$ values in the range of $10^{-7}$ to $10^{-9}$. The latest IUPAC panel recommends $\gamma_{ss}(BET) = 9×10^{-9}$ with a $\Delta\log\gamma_{ss} = 1$, i.e. an order of magnitude span around the mean. In **Figure S8**, the $\gamma_{ss}(BET)$ values obtained in the same concentration range (i.e. $10^{13}$ to $10^{14}$ molecules $cm^{-3}$) using the Knudsen cell are compared with literature values obtained for the relevant mineral substrates and the IUPAC recommended value. Our values fall within the recommended range.

Despite the extensive investigation of $NO_2$ uptake on mineral substrates, there has been no systematic study of the dependence of $\gamma_{ss}(BET)$ on gas-phase concentration. To explore this relationship and cross-validate our results, additional measurements were performed using the U-shaped reactor, as this setup allows better resolution of slow uptake processes, particularly for $\gamma_{ss}(BET) < 10^{-7}$. In **Figure 3**, $\gamma_{ss}(BET)$ as a function of $NO_2$ concentration for quartz and alumina is presented, measured with both experimental configurations, with good agreement observed between the two setups. Similar values were measured for both minerals, with a clear trend of increasing uptake with decreasing $NO_2$ concentration, and a power-law dependence on concentration similar to that observed for HCl. Further information on the concentration dependence of $\gamma_{ss}(BET)$ for $NO_2$ and how it compares with literature is presented in Supporting Information section **S.3.3**.

The adsorption isotherms at room temperature were determined for the four mineral surrogates using the Knudsen cell reactor. The results are presented in **Figure S9**. Overall, the total uptake of $NO_2$ was at least two orders of magnitude lower than that of HCl, in the range of $10^{10}$ to $10^{12}$ molecules $cm^{-2}$, approaching the detection limit of the technique at low $NO_2$ concentrations. This behavior is likely related to the nature of the surface sites targeted by $NO_2$. As a weak oxidizer, $NO_2$ has a specific affinity for strongly reducing surface sites, which are presumably much less abundant than the basic sites targeted by HCl. Among the studied materials, quartz exhibited the highest uptake, with values approximately ten times greater than those of amorphous silica. The measured $N_{ss}$ values followed a Langmuir-type adsorption behavior, characterized by an initial increase in uptake with concentration, followed by saturation at higher concentrations. It should be emphasized that at low $NO_2$ concentrations ($<1\times10^{13}$ molecules cm$^{-3}$) the determination of surface coverage is associated with relatively high uncertainties (~50%), due to the weak uptake of $NO_2$ on the mineral surrogates.

The Langmuir parameters determined from the isotherm fitting are summarized in **Table 3**. Quartz exhibited the highest equilibrium constant, with a $K_{Lin}$ value approximately 100 times greater than that of amorphous silica, the latter displaying the lowest partitioning coefficient among all samples. For calcite and alumina, $K_{Lin}$ values ranged between two to seven times higher than that of amorphous silica, confirming that amorphous silica possesses the lowest $NO_2$ uptake capacity among the mineral surfaces tested.

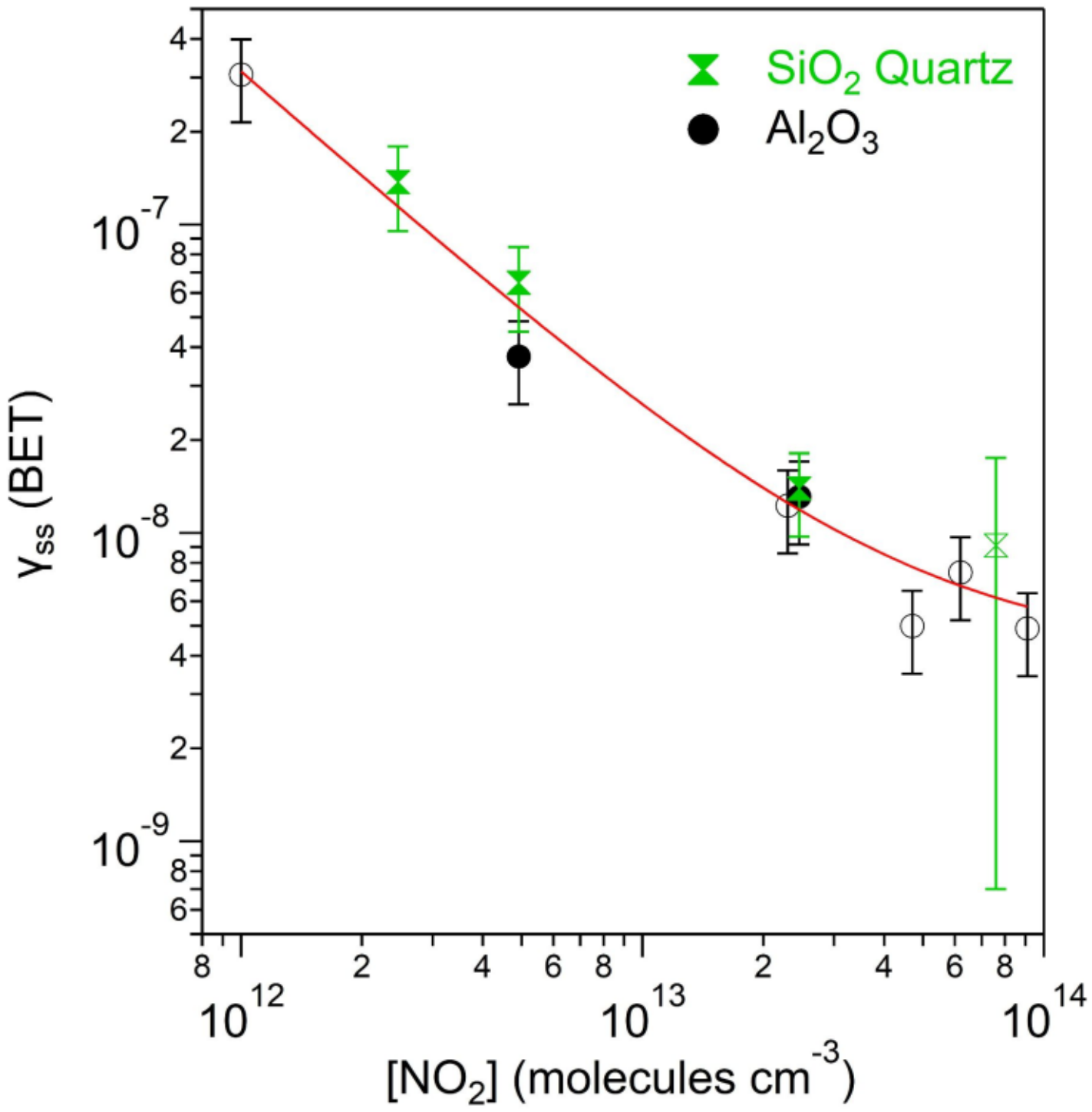

**Figure 3.** Steady state uptake coefficient of $NO_2$ on $SiO_2$ quartz and $Al_2O_3$ at room temperature. Filled symbols correspond to measurements performed with the U-shaped reactor while empty symbols are experiments performed with the Knudsen cell. Considering the close values of the uptake coefficients observed for both alumina and quartz, together with their similar trends, a common fit of the combined experimental data is presented.

# 4. Comparative analysis and SAI implications

The complete quantitative implications of our findings on SAI chemical impacts require detailed modeling beyond the scope of this study. As an initial assessment, we can estimate the direct impact of $NO_2$ uptake, and present a comparative analysis of the HCl uptake on the different minerals based on previous modeling studies.

During daytime, $NO_2$ concentrations in the lower and middle stratosphere are governed by rapid photochemical cycling, with photolysis as the dominant sink. Its daytime lifetime is typically on the order of minutes[48]. During nighttime, photolysis ceases and $NO_2$ is redistributed through reservoir formation, particularly $N_2O_5$ production followed by heterogeneous hydrolysis on ambient sulfate particle surfaces. To evaluate whether direct uptake of $NO_2$ on solid SAI particles could compete with these pathways, a representative aerosol loading of 5 Tg $yr^{-1}$ is considered (based on reference scenarios discussed by Vattioni et al.,)[26]. The first-order heterogeneous loss constant is given by

$$k_{het} = \frac{\bar{c}}{4} \times \gamma \times A' \qquad \text{Eq. 5}$$

where $\bar{c}$ is the mean molecular speed, $\gamma$ is the uptake coefficient, and $A'$ is the aerosol surface-area density. Using a conservative upper-limit value of $\gamma_{ss}(BET)$ = $5\times10^{-4}$ extrapolated from our measurements at stratospherically relevant $NO_2$ concentrations ($1\times10^{9}$ molecules $cm^{-3}$) and a particle specific surface area on the order of 5 $m^2$ $g^{-1}$ (for an optimally sized SAI particle of ~0.5 microns in diameter) yields a lower-stratospheric surface-area density of ~10-20 $\mu m^2$ $cm^{-3}$ and a heterogeneous removal timescale of ~5 to 10 months. Even with an extreme 100-fold larger surface-area density, the lifetime would still be several days, far longer than the photochemical lifetime of $NO_2$. Direct heterogeneous uptake of $NO_2$ on SAI particles is therefore expected to be negligible, with stratospheric impacts more likely arising indirectly through processing or physical removal of reservoir species such as $HNO_3$, $N_2O_5$ and $ClONO_2$ rather than through direct $NO_2$ loss on particles.

More generally, significant chemical impacts on the stratosphere, and particularly enhanced ozone depletion, are far more likely if catalytic cycles are involved. For a relatively non-reactive surface, the prominent reaction of interest is the heterogeneous reaction between $ClONO_2$ and HCl, which will release $Cl_2$ and eventually catalytically destroy ozone. Previous studies with surfaces of ice, nitric acid trihydrate and alumina have indicated that the underlying mechanism includes the adsorption of HCl and its inclusion in a reactive form on the surface; thus the parameter of interest for each surface is the reactive uptake coefficient for $ClONO_2$ in the presence of a given gas-phase concentration of HCl, which we denote $\gamma_{r,\mathrm{ClONO2}}$. For alumina, this coefficient has been measured[39] and found to be $\gamma_{r,\mathrm{ClONO2}}$ = 0.02 in the presence of 1 ppb of HCl at 50 hPa (corresponding to an altitude of ~20 km). Based on the modeling work of Vattioni[26], who used this value in conjunction with a 5 Tg $yr^{-1}$ injection scenario of alumina under 2020 halogen contents, a depletion of 1 to 10% of the total ozone column (TOC) could be expected, depending on the specific adsorption model assumed for extrapolation to relevant concentrations (e.g., dissociative vs non-dissociative adsorption). Additional calculations detailed in the supplementary information of that study, corresponding to the different adsorption models and assuming a range of uptake coefficients spanning almost two orders of magnitude (0.0003 to 0.019), predicted a corresponding trend in TOC depletion, i.e., $\gamma_{r,\mathrm{ClONO2}}$ is a first-order parameter, and the ozone impact is approximately proportional to it.

Similar measurements and modeling efforts should be conducted for any proposed surface, using relevant concentrations and temperatures. Absent the direct measurements needed to inform these models, we provide here a rough estimate of the relative reactivity of the surfaces based on the HCl adsorption isotherms. For low concentrations of HCl and $ClONO_2$ typical of the stratosphere, and neglecting any adsorption of $ClONO_2$ or direct reactions between this molecule and unoccupied surface sites, the Langmuir-Hinshelwood model indicates that the heterogeneous reaction rate will be limited by the surface occupancy and reactivity of adsorbed HCl. If the reactivity of an ambient $ClONO_2$ molecule with an adsorbed HCl molecule on the different surfaces were similar, and assuming all other parameters being equal, the surface coverage will directly define the reactive uptake coefficient. For all four surfaces, the coverage at room temperature but with stratospheric HCl concentrations is deep in the unsaturated regime, and, based on our measurements, the ratio of $N_{ss}$ is approximately 3400 : 1600 : 350 : 1 for calcite : alumina : quartz : amorphous silica, respectively. Therefore, under a similar injection scenario and the previous assumptions, and assuming the ratio of surfaces coverages and reactivities remains at lower temperatures, extrapolating from Molina's measurements for alumina we could expect values of $\gamma_{r,\mathrm{ClONO2}} \sim 0.04$, 0.004 and $10^{-5}$ for calcite, quartz and amorphous silica respectively. Thus, under the 5 Tg $yr^{-1}$ scenario discussed by Vattioni, amorphous silica would deplete <<0.1% of the TOC, compared to several percent or more if calcite were used. For reference, the natural inter-annual variability of the near-global TOC is on the order of 1%[67].

A few important assumptions should be stressed here. First, Vattioni's model included partial wetting of the alumina surface by sulfuric acid due to condensation; this, as well as potential coagulation with background sulfate aerosols, and, more generally, any surface aging processes, must be correctly accounted for in a complete calculation and comparison of alternative materials. Furthermore, as discussed and modeled by Vattioni, stratospherically abundant nitric acid may compete with HCl for adsorption sites and substantially lower the uptake coefficient[42]; the relative effect of this on different surfaces may vary. Additionally, the optical properties of different particles may change their chemical impact both directly and indirectly. For example, the net scattering efficiency of optimally-sized alumina particles is higher than that of silica, and thus less mass will be required for a given radiative cooling. The increased stratospheric heating for silica due to its infrared absorption in the atmospheric window may induce additional side effects which have been studied for sulfates, including larger impacts on ozone, circulation and precipitation[20,29,68–70]. The specific form of amorphous silica used here, fumed silica, is a low-density material, and the ratio of its scattering cross-section to BET surface area will be especially low. Finally, the injected particle size distribution and its evolution over time, e.g., coagulation between dispersed particles, will impact the trajectories, surface area and radiative effects of the particles, and these may well depend on the specific aerosol characteristics (i.e. the particle sizes and sticking coefficients).

All these assumptions and caveats must be accounted for before any definitive conclusions can be made on the relative impacts of the minerals studied here, and any suggestions of safety for SAI must include additional considerations[2], e.g. biosafety of the aerosols, or their impacts on clouds[71]. However, the marked differences in surface coverage in these sample materials - over three orders of magnitude, including two orders of magnitude between two forms of chemically “identical” silica - indicate that there is a potential for substantial mitigation of adverse effects through the choice or fabrication of the material surfaces. In particular, we suggest that an improved particle may be designed by combining the dense structure of the crystalline surrogates, ideally, with the bulk having high transparency in the atmospheric window, with the surface properties of the amorphous silica particles.

In all the materials studied here, the primary adsorption sites are expected to be hydroxyl terminations. However, the density of these surface sites, as well as their chemical nature, may vary. We suggest that the main advantage of silica over calcite and alumina arises from the weakly acidic nature of its surface sites, which do not efficiently promote adsorption of trace acids such as HCl. In contrast, the surfaces of calcite and alumina are strongly and weakly basic, respectively. The differences observed between crystalline and amorphous silica may stem from a difference in the density or accessibility of its surface silanol sites. For example, amorphous silica is prone to dehydroxylation (e.g., if heated during fabrication), leading to the formation of siloxane bridges in place of adjacent silanols[72]. While a more definitive investigation of the specific surface sites of each sample would be instructive, the marked difference already observed between crystalline and amorphous silica is in itself a strong motivation for considering surface treatments that would reduce the stratospheric reactivity of any proposed material.

The application of surface treatments to improve the chemical characteristics of candidate materials opens up options for minimizing additional risks and uncertainties. For example, both alumina[26] and silica particles[73] introduced to the stratosphere would acquire at least partial sulfuric acid or sulfate coatings. This transformation could alter the optical properties of the particles, as well as their sedimentation velocity and thus their lifetime. Furthermore, the repartitioning of the stratospheric sulfate content from the smaller, naturally occurring sulfate aerosols to sulfate shells on larger injected particles would likely increase the total sulfate surface area, leading to increased ozone depletion. By applying a hydrophobic surface treatment, the potential for such wetting could be minimized or even avoided completely. Such a surface would likely also be beneficial for reducing the potential to uptake polar molecules such as HCl.

# 5. Conclusions

We have presented a comprehensive experimental investigation of the heterogeneous uptake of three stratospheric trace gases, $HNO_3$, HCl, and $NO_2$, on four solid mineral surfaces proposed as candidate materials for stratospheric aerosol injection: calcite ($CaCO_3$), alumina ($\alpha$-$Al_2O_3$), crystalline silica (quartz), and amorphous silica (fumed silica). Measurements were performed using complementary Knudsen cell and U-shaped flow-through reactor techniques over wide concentration ranges, enabling reliable extrapolation of kinetic parameters to stratospherically relevant conditions.

Uptake coefficients of $HNO_3$ and $NO_2$ showed excellent agreement with literature and served to cross-validate the two experimental techniques. $HNO_3$ measurements highlighted the importance of signal modulation when measuring sticky species. $NO_2$ uptake was found to be irreversible on all surfaces, with low initial uptake coefficients suggesting that $NO_2$ targets a limited number of specific reactive surface sites on all four surfaces. Extrapolation to stratospherically relevant $NO_2$ concentrations suggests heterogeneous removal timescales of several months even under substantial aerosol loading, indicating that direct $NO_2$ uptake on SAI particles is unlikely to constitute a significant stratospheric perturbation.

For HCl, strong uptake was observed on all four mineral surfaces. The steady-state uptake coefficients exhibited a clear and strong concentration dependence across five orders of magnitude in concentration, consistent with surface site limited, non-first order kinetics well described by Langmuir adsorption behavior. Analysis of the reactive fraction of HCl uptake on calcite indicated that at stratospherically relevant concentrations the efficacy of a chlorine sequestration mechanism will be limited. The dominant fraction of HCl molecules taken up on all the surfaces will thus be adsorbed without reacting, and adsorption isotherms revealed that the HCl partitioning coefficient of calcite is more than $10^4$ times greater than that of amorphous silica, and approximately twice that of alumina, reflecting the dominant role of surface acid/base characteristics in governing uptake. Using the HCl adsorption isotherms as a proxy for the reactive uptake coefficient of $ClONO_2$ in the presence of adsorbed HCl, at stratospheric concentrations we find marked, orders of magnitude differences in surface coverage among the materials, which could carry direct implications for the relative ozone impacts of these candidate materials.

Taken together, our results highlight that the chemical reactivity of candidate SAI materials is highly sensitive to both chemical composition and surface microstructure. The substantial difference observed between crystalline and amorphous silica, despite their identical bulk chemical composition, underscores the potential for designing engineered particle surfaces with reduced heterogeneous reactivity. We suggest that particles combining a dense, infrared-transparent core with a low-reactivity shell or surface treatment may represent a promising direction for minimizing the chemical side-effects of SAI, while also reducing the susceptibility of interactions with background sulfates. These considerations, alongside biosafety, optical properties, and particle size evolution, should inform the choice or design of candidate particles tailored for SAI.

# Acknowledgments

The authors thank Michel Rossi of EPFL for fruitful discussions on the kinetic measurements and interpretation of experimental observations.

# Disclosure

Y.S., T.K. and G.S.R. are affiliated with Stardust Labs Ltd., a sunlight reflection technology development company.

# Supporting Information material

## Uptake of stratospheric species on minerals proposed for stratospheric aerosol injection

Anais Lostier,[1] Yair Segev,[2,*] Tzemah Kislev,[2] Gal Schwartz Roitman,[2] Nadine Locoge,[1]
Manolis N. Romanias.[1,*]

1: IMT Nord Europe, Institut Mines-Télécom, Univ. Lille, Centre for Energy and Environment, F-59000 Lille, France,

2: Stardust Labs Ltd., Ness Tziona, Israel 7403638

# S.1 Gas synthesis

## S1.1 Nitric acid ($HNO_3$) preparation

Nitric acid ($HNO_3$) gas was prepared using a liquid-phase mixture of concentrated sulfuric acid ($H_2SO_4$, 95%, VWR) and $HNO_3$, maintained at 8°C to reduce the vapor pressure of $HNO_3$ and control its release. Due to the reactive and unstable nature of $HNO_3$, gas-phase experiments were performed using only the vapor emanating from this solution. The presence of water vapor was monitored throughout the experiments and remained below the detection limit and thus $HNO_3$ purity was always higher than 99%. The stability of the gas was monitored with the modulated quadrupole mass spectrometer of the Knudsen cell by tracking the *m*/*z* 46/30 signal ratio, which is 1.8. This ratio is consistent with the value reported by Santschi et al. (2006), who observed a ratio of 1.7, using the same experimental equipment[1].

## S1.2 Hydrogen chloride (HCl) synthesis

Hydrogen chloride (HCl**)** was synthesized in the laboratory by reacting potassium chloride (KCl) with a few drops of concentrated sulfuric acid ($H_2SO_4$, 95%, VWR), producing gaseous HCl and a solid byproduct ($KHSO_4$). The HCl gas was then condensed into a pre-cooled cell maintained at 77 K. After condensation, a purification step was carried out by pumping down under vacuum to remove any volatile impurities. The cell was subsequently connected to the vacuum line and gradually warmed up, allowing the controlled release of purified HCl gas. The purity was assessed by quadrupole mass spectrometry, which confirmed the absence of water vapor and a ratio of *m*/*z* = 36 to *m*/*z* = 38 signals consistent with HCl purity greater than 99%. Neat HCl stored in a glass bulb of 4 L was used to supply the Knudsen cell reactor for the uptake coefficient measurements in the range of $2\times10^{11}$ to $9\times10^{13}$ molecules $cm^{-3}$. Additional experiments were conducted at low HCl concentrations (~$7\times10^{9}$ molecules $cm^{-3}$), using a mixture of HCl diluted in helium at a concentration of 4.8%. Notably, for HCl, the detection limit, defined as three times the signal-to-noise ratio, is on the order of $2\times10^{9}$ molecules $cm^{-3}$. This enhanced sensitivity is primarily attributed to the effective modulation of the molecular beam in the mass spectrometer setup.

## S1.3 Nitrogen dioxide ($NO_2$) preparation

Nitrogen dioxide ($NO_2$) was introduced into a glass bulb from an $NO_2$ gas cylinder (Alpha Gaz, > 99%). The bulb was filled to approximately 10 Torr and shielded from light to prevent photolysis. The gas was used directly for experiments in the Knudsen cell. The stability of the gas was monitored with the modulated quadrupole mass spectrometer of the Knudsen cell by tracking the ratio of *m*/*z* =46 to *m*/*z* = 30 signals, which is 0.3. For the experiments carried out in the U-shaped reactor, $NO_2$ was supplied from a canister filled with a mixture of $NO_2$ (from the cylinder) diluted in helium. The concentration in the canister was verified using a NO/$NO_2$ analyzer (Teledyne 200 UP, with photolytic convertor), and was approximately 30 ppmV. The concentration of $NO_2$ introduced into the U-reactor was then precisely controlled by dilution using two mass flow controllers: one for $NO_2$ and one for zero air.

# S.2 Experimental methods

## S.2.1 The Knudsen cell reactor

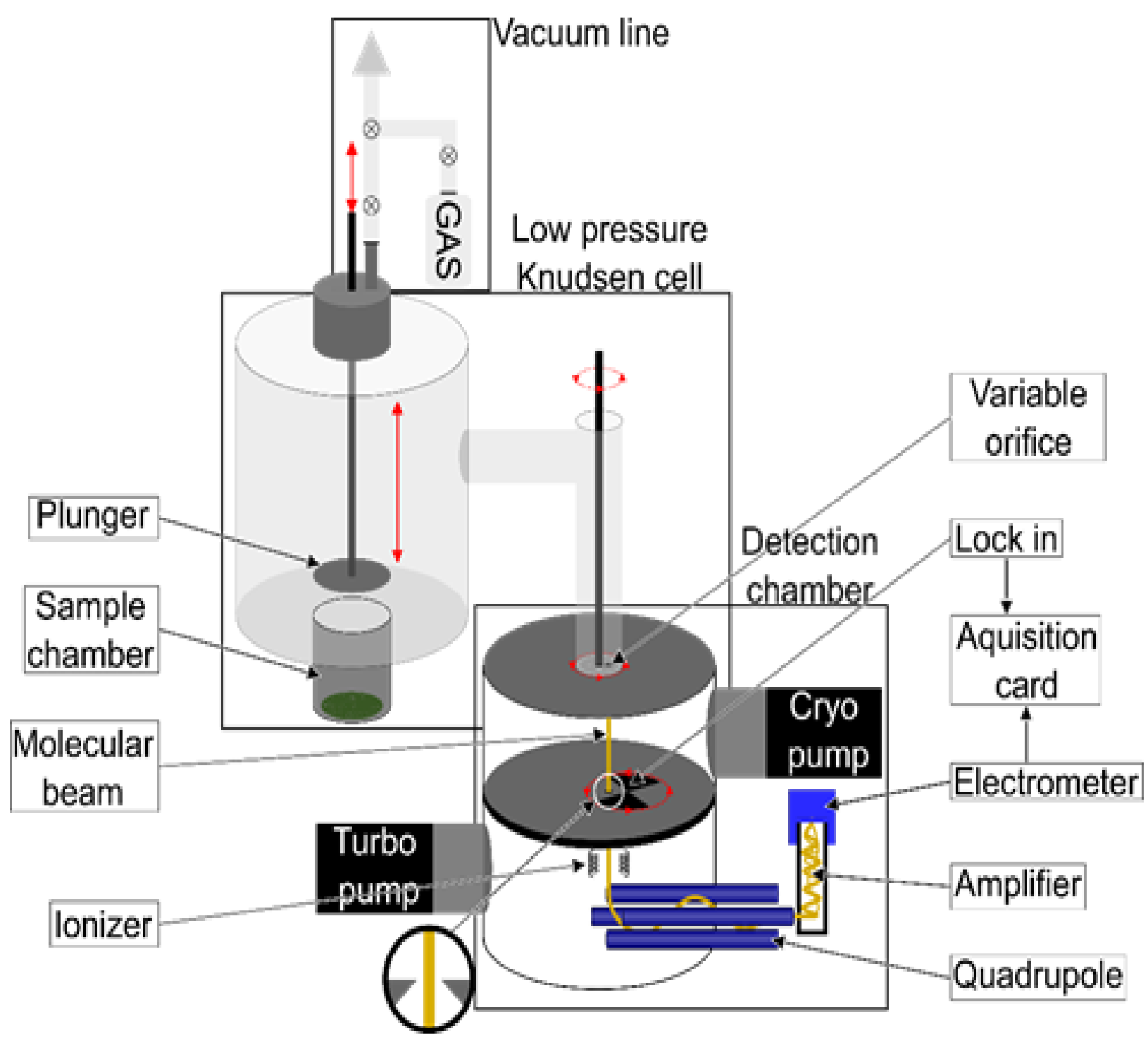

**Figure S1.** General scheme of the Knudsen cell, adapted from Pascaud et al.,[2].

#### S.2.1.1 Importance of modulation for reliable kinetic measurements

Molecules that escape from the Knudsen cell reactor enter a differentially pumped mass spectrometer chamber where they are modulated at 140 Hz by a chopper and detected by a quadrupole mass spectrometer (QMS, Balzers QMG 421, 70 eV). The chopper frequency is synchronized with a lock-in amplifier (SR830 DSP), ensuring only molecules exiting the cell and forming the thermal molecular beam are detected (phase-sensitive detection). The modulation approach is essential for accurately quantifying the uptake of sticky species such as HCl and $HNO_3$, as it minimizes wall memory effects that would otherwise lead to significant underestimation of uptake coefficients. Representative uptake profiles are shown in **Figure S2**, comparing the raw mass spectrometer (MS) signal and the lock-in amplifier signal (both normalized for clarity) for two compounds: a non-sticky species (glyoxal) and a highly interactive species (HCl).

For glyoxal, the unmodulated MS response and the lock-in output signal fully overlap, indicating that the detected intensity arises exclusively from molecules exiting the reactor. This confirms that background contributions within the MS chamber are negligible and stable over time, and that the measurement reliably reflects the modulated molecular beam at 140 Hz. In contrast, for HCl, a pronounced discrepancy is observed between the unmodulated MS response and the lock-in output, particularly during the initial stages of surface exposure. Although the background is subtracted prior to HCl introduction, this correction assumes a stable baseline, which is not the case. Due to its strong surface affinity, HCl interacts with the internal surfaces of the stainless-steel chamber housing the mass spectrometer, leading to slow desorption and a time-dependent background that evolves during the experiment. As a result, when rapid uptake causes a sharp decrease in gas-phase concentration in the reactor, the MS response remains biased by this drifting background within the detection chamber, leading to an underestimation of the true uptake. Beam modulation at 140 Hz, combined with phase-sensitive detection using a lock-in amplifier tuned to the same frequency, selectively isolates the signal corresponding to molecules exiting the reactor. This approach effectively removes contributions from non-modulated, slowly varying background species, enabling accurate measurement of transient concentration changes. The advantage is most pronounced during the initial exposure phase, where concentration gradients are steep. As the surface approaches saturation, the discrepancy between the two signals decreases; however, the lock-in signal remains the more reliable representation.

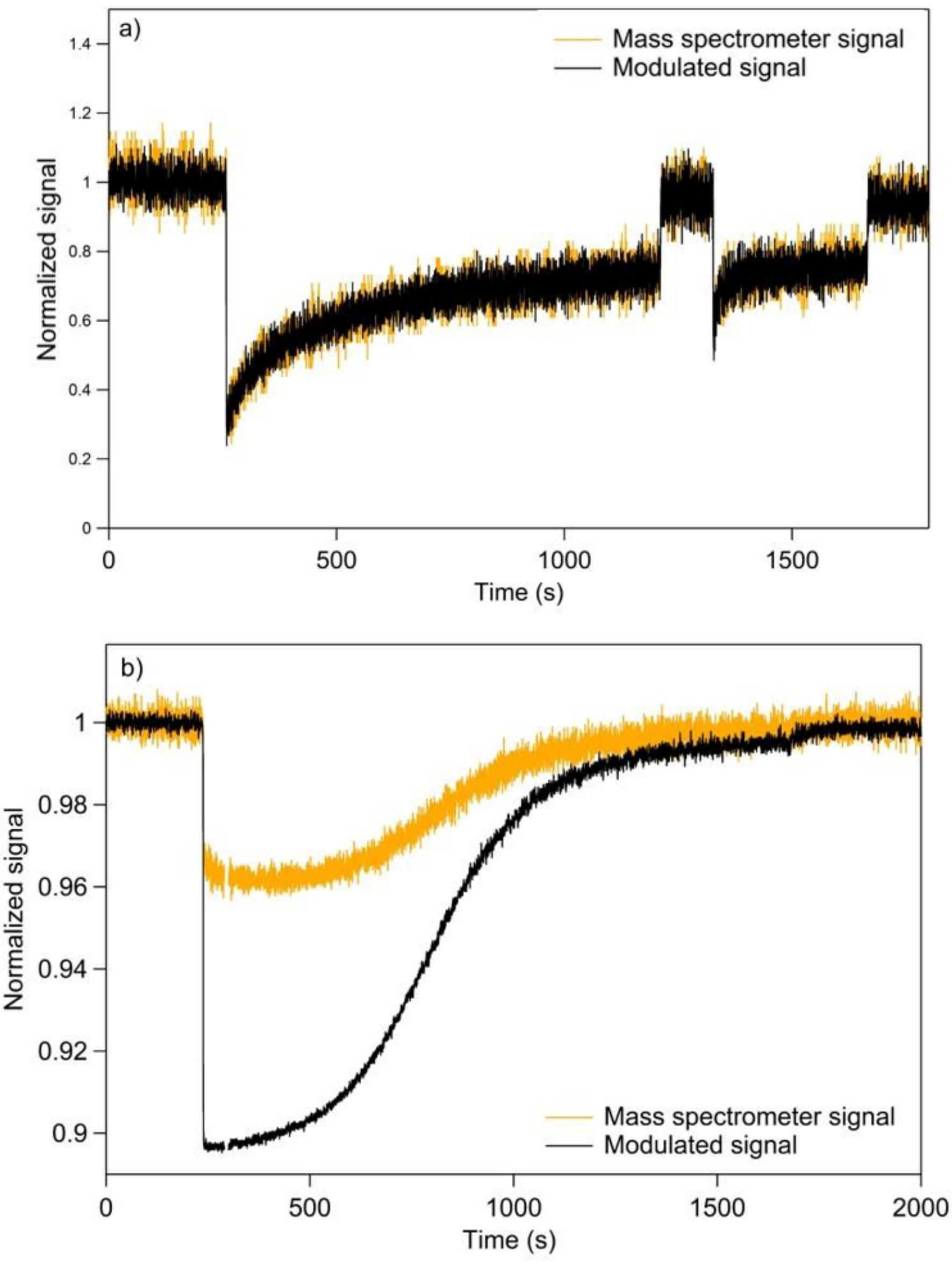

**Figure S2.** Typical normalized uptake profiles of a) methylglyoxal on Arizona dust, and b) HCl on calcite. The orange curves represent the MS signal without any modulation, and the black curves present the demodulated signal using the lock-in amplifier at 140 Hz.

Importantly, memory effects for sticky species are cumulative, and their impact increases with repeated measurements. In conclusion, the advantage of modulated molecular beam experiments has been noted for more than six decades in the literature, and is also discussed here. Uptake coefficients and adsorption parameters derived from unmodulated molecular beam experiments, particularly initial uptake values, should be interpreted with caution, as they may be systematically underestimated.

### S.2.1.2 Surface coverage integration and steady-state

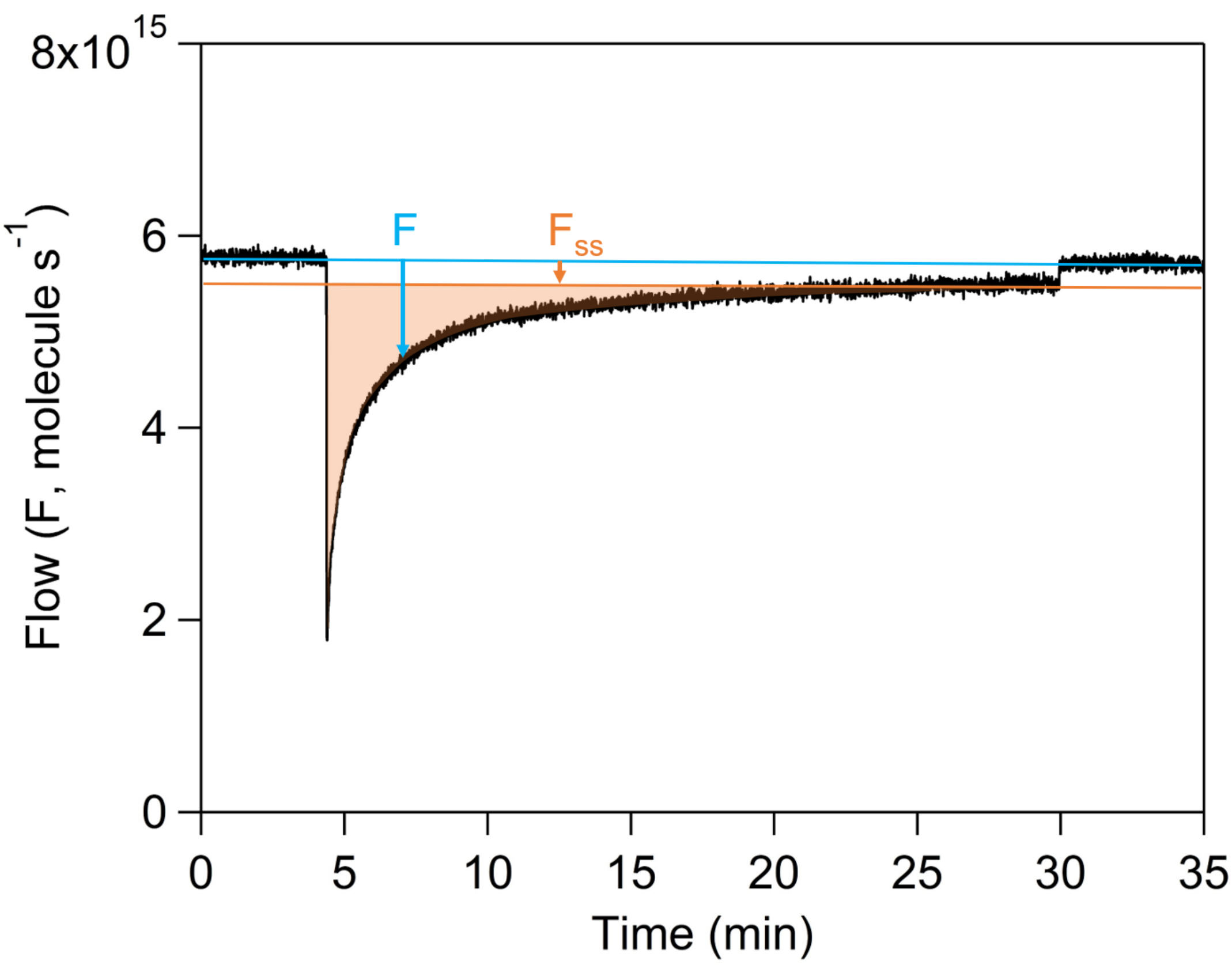

**Figure S3.** Example of a Knudsen cell reactor uptake measurement for HCl on $Al_2O_3$, performed with an escape orifice diameter of 8 mm, with sample exposure starting after about 4 minutes. The surface coverage $N_{ss}$ is calculated by integrating the orange region until reaching a steady state criterion. As the integration does not include the steady-state fraction of the uptake, $N_{ss}$ will not depend on prolonged exposure after reaching steady-state. This approach has been applied widely in the literature to quantify the uptake of a non-saturated process (i.e. one with a non-zero steady-state uptake)[3].

## S.2.2 The U-shaped reactor

A U-shaped flow reactor was used to determine steady-state $NO_2$ uptake coefficients on the mineral surfaces and to cross-validate these kinetics with Knudsen cell measurements. Dry zero air was used as the bath gas, with estimated water concentration levels of $< 4.8 \times 10^{13}$ molecules $cm^{-3}$ (as provided by the supplier). A schematic of the reactor is shown in **Figure S4**. It comprises two main sections: the upper section is a bypass line (volume: 15 $cm^3$) used to monitor $NO_2$ concentrations in the absence of a sample, while the lower section (also 15 $cm^3$) contains a fritted glass sample holder where a defined amount of sample is deposited. The two parts are connected via a pair of 3-way valves, allowing switching between bypass and exposure modes. Calibrated mass flow controllers (Bronkhorst) were used to introduce $NO_2$ and dilute it with zero air upstream of the reactor at a total flow rate of 300 sccm.

For each experiment, a known quantity of solid sample (~20 mg) is weighed and deposited into the U-shaped reactor. The sample is then flushed overnight with dry air to remove any pre-adsorbed species from the surface. Real-time monitoring of $NO_2$ was carried out using a selected-ion flow tube mass spectrometer, SIFT-MS Voice 200. This technique relies on the chemical ionization of analytes by three reagent ions: $H_3O^+$, $NO^+$, and $O_2^+$. $NO_2$ was detected using $O_2^+$ as the reagent ion, at mass 46 ($NO_2^+$).

In a typical experiment, $NO_2$ is first introduced into the bypass of the reactor, and its concentration is monitored using the SIFT-MS until it reaches a stable level. Once the signal is stabilized, the gas flow is directed through the U-shape reactor containing the solid sample. The sample is exposed to the gas until a steady-state signal is observed, indicating that the interaction between the gas and the surface has reached an equilibrium. Then, the gas flow is redirected back through the bypass to evaluate the nominal concentration of $NO_2$. The $NO_2$ flow is then stopped, maintaining the dry air flow, to record the baseline signal. Finally, the sample is flushed with zero air to record the desorbed fraction.

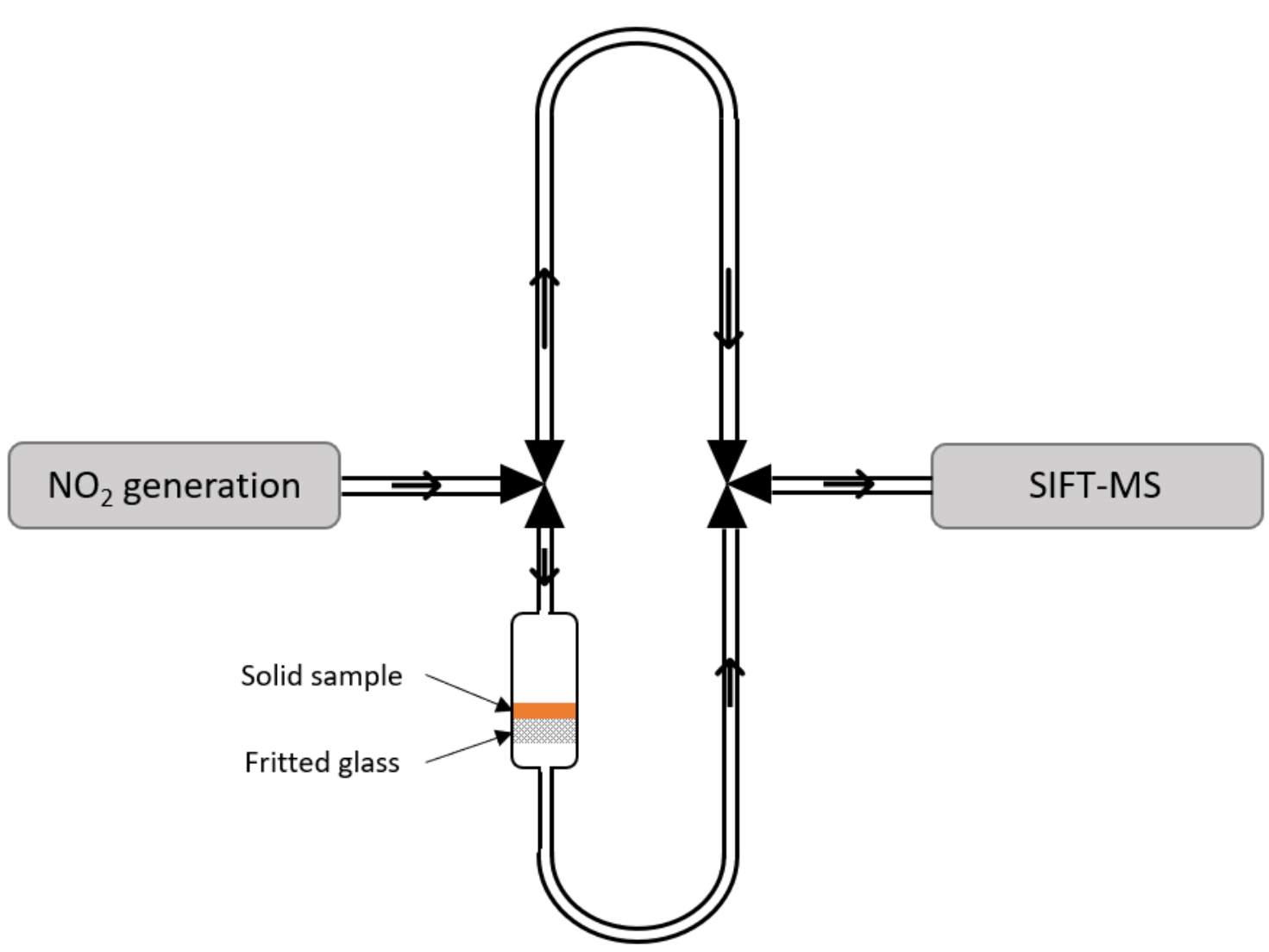

**Figure S4.** Schematic representation of the U-shaped reactor used to determine the steady state uptake coefficients of $NO_2$ on alumina and quartz.

## S.2.3 Calculating uncertainties

The uncertainties in $\gamma_0$, $\gamma_{ss}$ and $N_{ss}$ measurements were determined using the quadrature method of uncertainty propagation, taking into account systematic uncertainties. In particular, the uncertainty in the purity of the gas sources was on the order of 1%. For experiments in the Knudsen cell, the uncertainties considered include flow measurement accuracy (5%), signal precision (2σ) 2% in concentrations above $1\times10^{10}$ molecule $cm^{-3}$ and 5% at lower concentrations, errors in $k_{esc}$ measurements (10% with a 14 mm orifice, 1% for 1 mm). For U-shaped reactor experiments, the corresponding errors are 5% for the flow rate and 10% for signal precision (2σ) of the SIFT-MS. In both experimental systems, for $\gamma_{ss}$ and $N_{ss}$, additional sources of uncertainty include the specific surface area (2.4-12.5%) and sample mass (0.5%). For $N_{ss}$, uncertainty related to signal integration (10%) is also considered. The overall error including reproducibility (i.e. multiple experiments under the same conditions, which was always better than 5%), and the systematic uncertainties presented above, are: $\gamma_0$ ~13%; $\gamma_{ss}(BET)$ ~18%; and $N_{ss}$ ~20% for the Knudsen cell measurements. For the U-shape reactor, the uncertainty in $\gamma_{ss}(BET)$ is also ~18%.

# S.3 Results

## S.3.1 $HNO_3$ uptake on mineral surrogates

**Figure S5** presents the initial uptake coefficient of nitric acid on alumina and calcite for different studies in the literature, and for the current study. There is a good agreement between the literature values in the studied concentration range, except the Goodman et al. study and Vlasenko et al. Goodman et al.'s measurements were conducted using an unmodulated molecular beam mass spectrometer. Consequently, their data are likely affected by memory effects in the mass spectrometer's high-vacuum chamber, leading to an underestimation of uptake coefficients, as discussed in section **S.2.1.1**. In Vlasenko et al., which used an aerosol flow tube, the residence time was about 3 minutes, significantly longer than in Knudsen cell measurements that were on the order of milliseconds, likely underestimating the peak initial value. Excluding these studies, the concentration dependence of the initial uptake coefficient is well described by the following expression:

$$\gamma_0(HNO_3) = 6.0 \times 10^{-2} + (3.5) \times 10^5 [HNO_3]^{-0.6} \quad \text{Eq. S.1}$$

Although in the real troposphere or stratosphere $HNO_3$ concentrations are expected to be at lower levels, the above expression describes the general trend observed in the initial uptake coefficient values. Given the extensive prior research on $HNO_3$ uptake, this molecule is selected as a benchmark to evaluate the agreement of the Knudsen cell measurements against kinetic results found in the literature. Our results show excellent agreement with previous studies, demonstrating the reliability of our experimental setup.

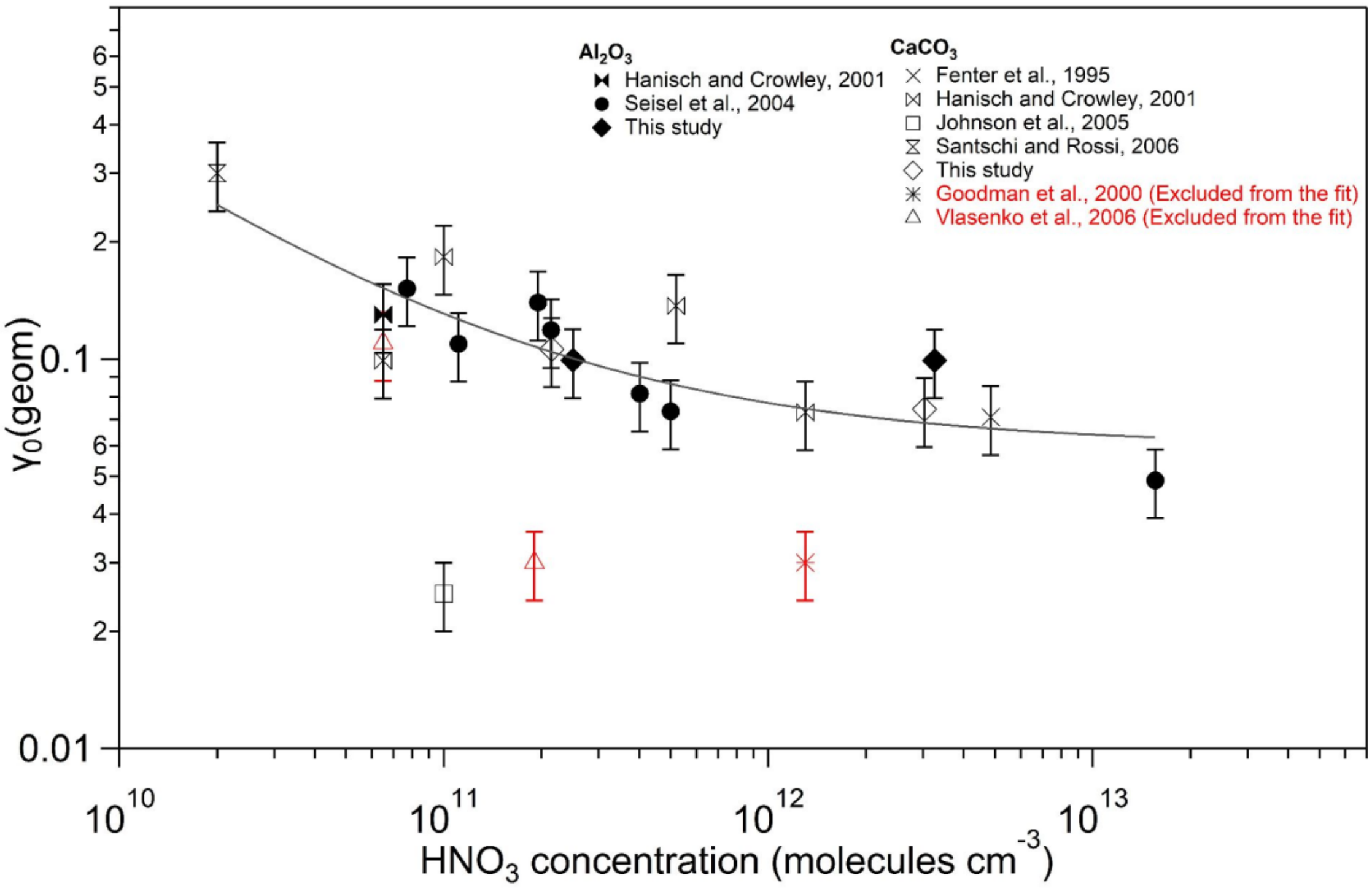

**Figure S5.** Initial uptake coefficients as a function of nitric acid concentration at room temperature, as reported in the literature and in the present study. Filled symbols correspond to measurements on $Al_2O_3$, all other symbols correspond to measurements on $CaCO_3$. Filled and empty diamond symbols correspond to measurements performed in this study. Results in red are excluded from the fit, as explained in the text. Literature values corresponds to: Fenter et al., 1995; Goodman et al., 2000; Hanisch and Crowley, 2001; Johnson et al., 2005; Santschi and Rossi, 2006; Seisel et al., 2004; Vlasenko et al., 2006[1,4–9].

# S.3.2 HCl uptake on mineral surrogates

### S.3.2.1 Steady state uptake coefficients and comparison with the literature

As shown in **Figure 1**, under stratospherically relevant concentrations, the values of $\gamma_{ss}$*(BET)* fall within the range of $10^{-3}$ for quartz and calcite and almost $10^{-2}$ for alumina, suggesting that these particles remove HC relatively efficiently. The steady state uptake of amorphous silica is around two orders of magnitude lower, indicating a significantly weaker interaction with HCl. At lower temperatures, HCl uptake is expected to increase, as reduced thermal energy leads to longer gas–surface contact time, which is exponentially dependent on temperature[10]. However, if the uptake is reactive, the formation of surface products may reduce overall uptake at lower temperatures, depending on the reaction mechanism. To obtain more realistic estimates and better assess the impacts of SAI, it is essential to investigate the effect of temperature on steady-state uptake coefficients.

The results of the current study can be compared with those obtained in the literature. Santschi et al. reported an initial uptake coefficient of HCl on highly oriented precipitated $CaCO_3$ of 0.13 following thermal treatment to remove strongly bound water. This value increased to 0.2 with elevated surface water content. In comparison, the initial uptake coefficients determined in the present study are higher by a factor of 2 to 4. This discrepancy is likely attributable to differences in the physical characteristics of the $CaCO_3$ particles used in the two studies, particularly their size and associated surface hydration. While our calcite nanoparticles were thoroughly pumped prior to exposure, reducing gas-phase water to background levels (<$10^{10}$ molecules $cm^{-3}$), strongly adsorbed water could not be entirely removed. Literature evidence indicates that smaller particles retain higher concentrations of surface OH groups due to their greater capacity to maintain water[11]. In our study, 90 nm calcite particles were employed, whereas Santschi et al. used particles approximately 2 μm in diameter[1]. The enhanced water retention of nanoparticles is thus a plausible explanation for the higher uptake observed. This interpretation is further supported by a comparison of typical HCl uptake profiles, such as those shown in Figure 10 of Santschi et al[1].

Huynh et al. investigated the initial uptake coefficients of HCl on $CaCO_3$ nanoparticles (200 nm diameter) at both 298 K and 207 K using an aerosol flow tube, as well as bulk samples analyzed via attenuated total reflectance infrared spectroscopy (ATR-IR)[12]. The uptake coefficients derived from the flow tube correspond to initial uptake values, given the short reaction times examined, approximately 2.3 seconds for the low-temperature experiments and 13 seconds at room temperature. A pronounced temperature dependence was reported, with the initial uptake coefficient increasing by a factor of six at 207 K (0.076 ± 0.009) compared to 298 K (0.0013 ± 0.001). While this difference was attributed to thermal effects, it is important to note that comparing initial uptake coefficients determined over different interaction durations may also introduce discrepancies (i.e., a longer duration may “smear” a high initial uptake peak), potentially leading to underestimation at room temperature (where the duration was longer) relative to the colder conditions in that study. In the present study, the initial uptake coefficient measured on $CaCO_3$ is approximately 33 times higher than that reported by Huynh et al., likely due to the use of geometric surface area in scaling the uptake coefficient in the Knudsen cell experiments. Although applying the geometric surface area is valid in the initial uptake regime (in the millisecond range), where only the outermost layers of the sample are accessible to the gas within the sub-second timescale, it results in an upper-limit estimate for the uptake coefficient.

Concerning steady-state uptake coefficients, at low temperatures (200–215 K), Dai et al. investigated the uptake of HCl on calcite particles using a low-pressure coated-wall flow tube and a flask-based system[13]. Experiments were conducted over a broad concentration range ($10^{11}$ to $10^{17}$ molecules $cm^{-3}$). The resulting uptake coefficients, presented in **Figure 1**, exhibit concentration-dependent trends similar to those observed in the present study. The reported values are slightly higher than those measured in the current study, which most likely can be attributed to the lower experimental temperature. However, several methodological differences between the two studies should be mentioned. One is the surface exposure time the flow-tube experiment in Dai et al. employed 10-15 min exposure, compared with ~1 h or longer in the present study. Nevertheless, their uptake transients appear to reach a steady state within that interval. Additional contributors may include gas-phase mass-transfer limitations, differences in surface preparation (e.g., solvent-assisted coatings in the flow tube), and uncertainties in the estimation of the effective surface area available for reaction.

### S.3.2.2 Reactive uptake of HCl on $CaCO_3$

A typical uptake experiment of HCl uptake on $CaCO_3$ is presented in **Figure S6**. Once the surface is exposed, a rapid decrease in HCl concentration is noted due to its uptake. Over time, the HCl concentration in the gas phase gradually recovers, attributed to the depletion of reactive surface sites, while $CO_2$ formation stabilizes. A critical surface coverage threshold appears to exist, above which $CO_2$ formation becomes linearly dependent on HCl consumption. This relationship is illustrated in **Figure S7**, where the $CO_2$ concentration is plotted against the total amount of HCl molecules taken up ("consumed"), revealing a yield of 80% for this specific concentration of $4.1\times10^{13}$ molecules cm$^{-3}$.

**Table S1** summarizes three such measurements performed at different concentrations of HCl. The ratio of HCl molecules consumed and $CO_2$ molecules formed decreases with increasing HCl concentration. A similar trend is noted for the product yield ($2\Delta[CO_2]/\Delta[HCl]$), beyond the threshold where HCl uptake correlates linearly with $CO_2$ formation. These observations highlight the critical role of surface coverage in governing HCl reactivity. According to our observations, under stratospherically relevant concentrations, the reactive fraction of HCl can fall below 20%, indicating that a substantial portion of the adsorbed HCl remains unreacted and may be available to interact with species such as $ClONO_2$.

Consequently, the complete chemical conversion of $CaCO_3$ particles to $CaCl_2$, which has been proposed as a strategy to remove stratospheric chlorine, may not be achieved at low HCl concentrations[14]. Furthermore, although not commonly addressed in the SAI literature, the potential reactivity of $CaCl_2$ particles with $ClONO_2$ should not be dismissed[13]. Previous studies have demonstrated the heterogeneous reactivity of salt surfaces such as NaCl with $ClONO_2$, suggesting that similar reactions may also occur on $CaCl_2$ surfaces[15,16].

**Table S1.** Summary of results obtained for the reactive fraction of HCl on $CaCO_3$ at room temperature across various HCl concentrations.

| [HCl] (molecules cm$^{-3}$) | HCl consumed (molecules cm$^{-2}$) | $CO_2$ formed (molecules cm$^{-2}$) | Total reactive fraction (%)[a] | $CO_2$ yield after coverage threshold |
|---|---|---|---|---|
| $6.7\times10^{9}$ | $3.4\times10^{14}$ | $2.7\times10^{13}$ | 15.6 | 18.5% |
| $2.6\times10^{11}$ | $2.7\times10^{14}$ | $7.4\times10^{13}$ | 54.4 | 60.4% |
| $4.1\times10^{13}$ | $2.3\times10^{14}$ | $9.4\times10^{13}$ | 82.2 | 90% |

[a]: determined by integrating the $CO_2$ formation and HCl uptake peaks, considering also the stoichiometry of the reaction

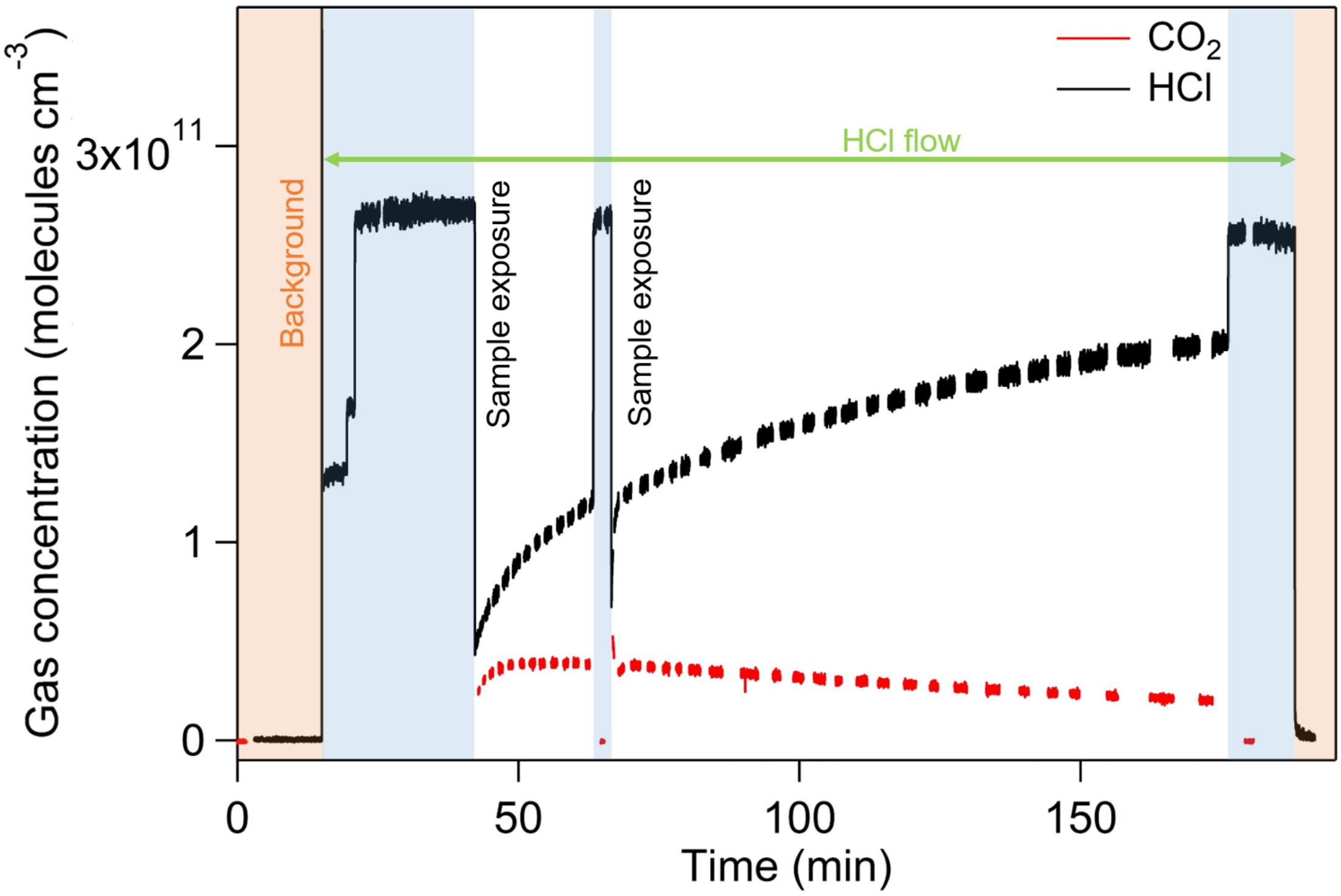

**Figure S6.** Typical uptake profile of HCl taken up on $CaCO_3$, and the simultaneous formation of $CO_2$ as a product of the reaction. The experimental conditions were: mass 141.9 mg, [HCl] = $2.7 \times 10^{11}$ molecules cm$^{-3}$, orifice diameter 14 mm, HCl and $CO_2$ monitored in *m*/*z* = 36 and 44, respectively.

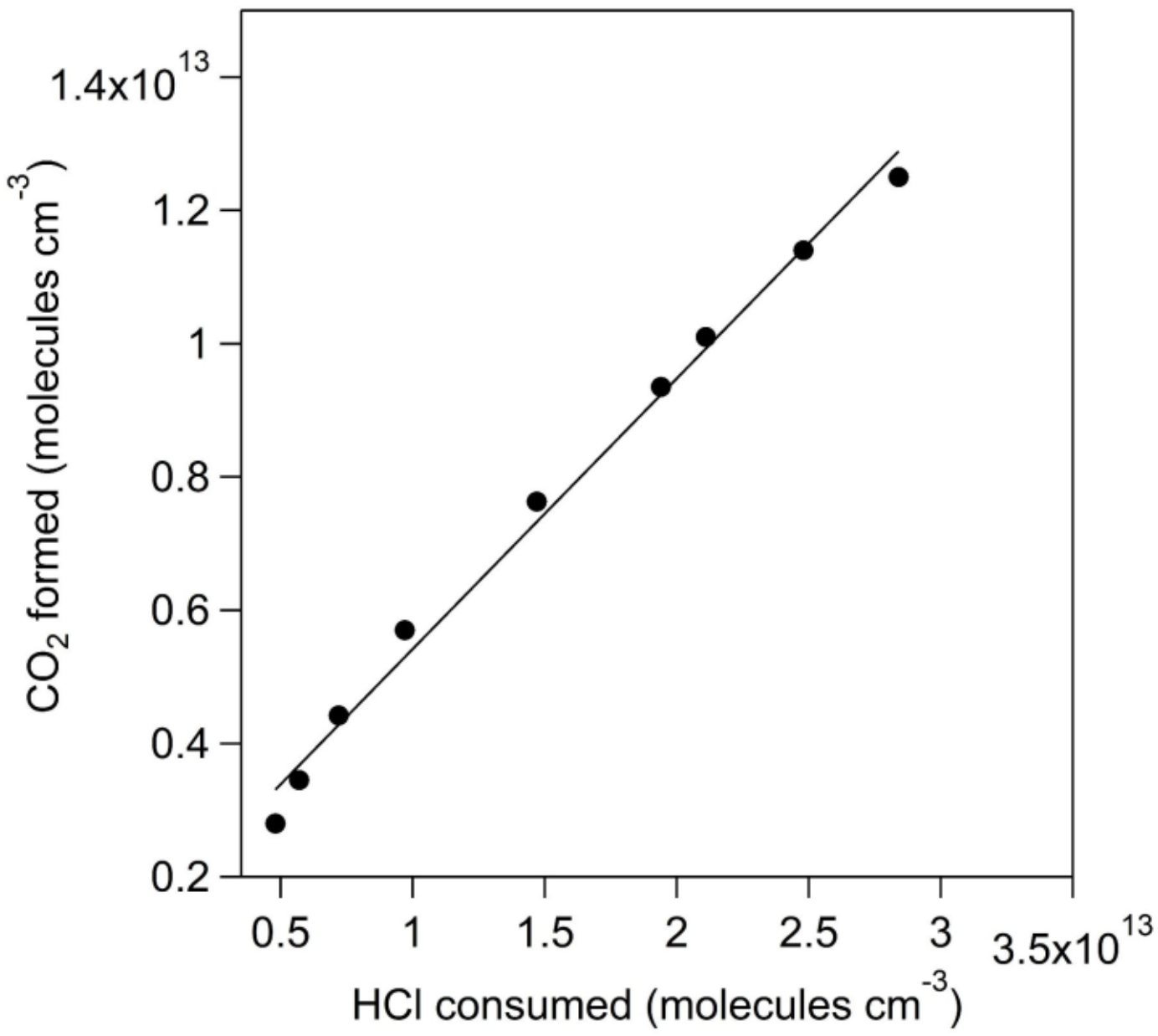

**Figure S7:** $CO_2$ formed as a function of HCl consumed at HCl concentration of $4.1 \times 10^{13}$ molecules cm$^{-3}$, measured in the Knudsen cell at T = 296 K. The observed trend is $[CO_2]_{formed}=(1.4\pm0.2)10^{12} + (0.41\pm0.02)[HCl]_{consumed}$, with $r^2 > 0.99$. The figure displays the linear relationship between $CO_2$ molecules formed and HCl molecules consumed after reaching a critical surface coverage level where a linear relationship links $CO_2$ formation and HCl consumption.

## S.3.3 $NO_2$ uptake on mineral surrogates

The $NO_2$ steady state uptake coefficients were found to be in excellent agreement with the literature (**Figure S8)**, but a strong concentration dependence was noticed (**Figure 3)**. Experimental results for alumina and quartz were combined and fitted with a single power-law expression since their uptake coefficients were similar. When extrapolating the concentration dependence to stratospherically relevant $NO_2$ levels (i.e., ~ $1.3\times10^9$ molecules $cm^{-3}$), the estimated $\gamma_{ss}(BET)$ value for these minerals is $5.7\times10^{-4}$. It should be noted that, for $NO_2$, extrapolation to stratospherically relevant concentrations remains uncertain, as the lowest concentration investigated is approximately three orders of magnitude higher than stratospheric levels. Nevertheless, the present dataset clearly demonstrates a concentration dependence of the process.

In the literature, Börensen et al. investigated the reactive uptake of $NO_2$ on $\gamma$-$Al_2O_3$ using *in situ* diffuse reflectance infrared Fourier transform spectroscopy (DRIFTS), reporting uptake coefficients in the range of $10^{-8}$ to $10^{-9}$ and observing an increase in the reactive uptake coefficient with increasing $NO_2$ concentration, a trend opposite of that observed in our study[17]. However, a direct comparison between the two studies is not straightforward due to significant differences in experimental approach and surface preparation. Börensen et al.'s method allowed them to monitor only the reactive fraction of the uptake, based on surface modifications, whereas our measurements account for the total $NO_2$ uptake, which includes both physisorption (non-reactive) and chemisorption (reactive) components. As such, while the reactive uptake may increase with concentration, the total uptake can still decrease, reconciling the apparent contradiction between the two studies. Moreover, the γ-alumina used in Börensen et al. underwent extensive pretreatment, including mechanical grinding to expose new reactive surface sites and thermal treatment at 573 K under vacuum, which likely led to partial dehydration and alteration of surface hydroxyl functional groups. These modifications can significantly alter surface reactivity compared to our study, which used $\alpha$-$Al_2O_3$ without such pretreatment. Overall, due to the distinct experimental protocols and the nature of the measured uptake (reactive vs. total), the results of Börensen et al. cannot be directly compared with ours.

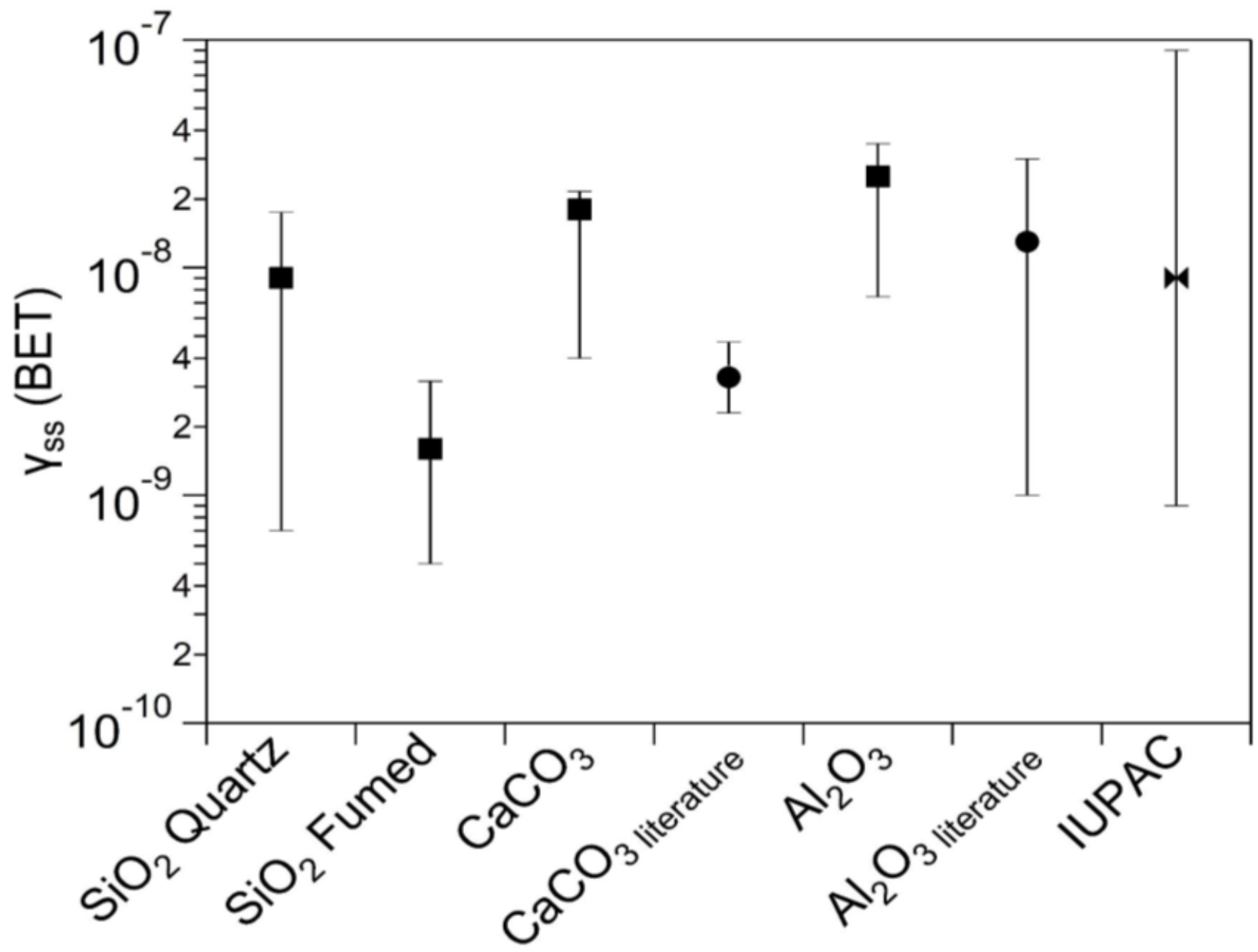

**Figure S8**. Steady state uptake coefficient of $NO_2$ on our samples, compared with literature data and the IUPAC recommended value, obtained over a concentration range of $10^{13}$ to $10^{14}$ molecules $cm^{-3}$ [18]. Literature measurements were performed by DRIFTS at room temperature[17,19]. Measurements of such low uptake coefficients approach the detection limit of the Knudsen cell technique, leading to relatively high uncertainty, estimated to be in the range of 30% for some minerals. In the case of quartz, the reported value should be considered an upper limit (determined as twice the precision of the measurement) for the investigated concentration range.

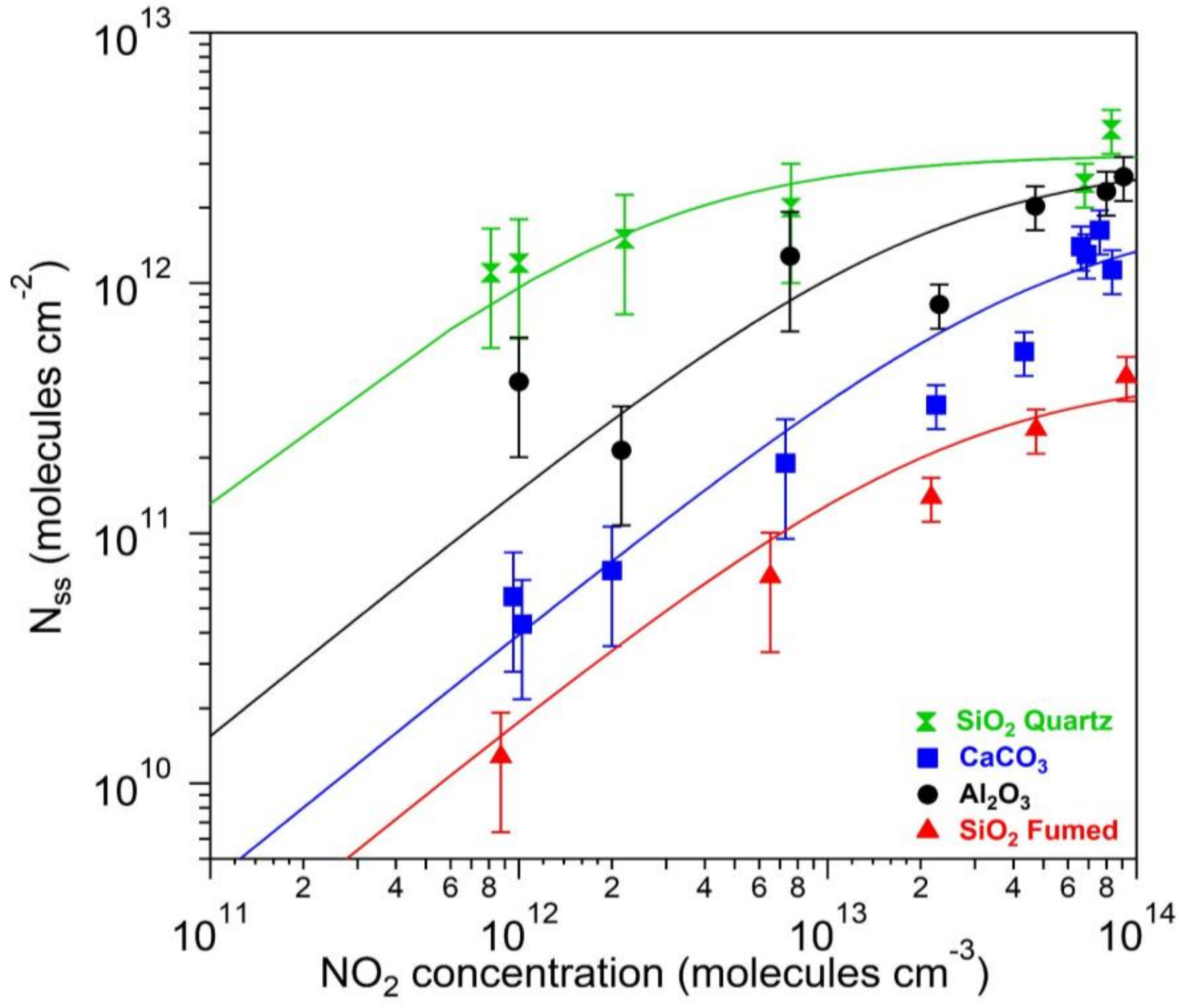

**Figure S9.** Adsorption isotherms of $NO_2$ on various mineral surrogates, determined at room temperature. Experimental results are fitted using the Langmuir adsorption isotherm model. Note that these values do not account for the reactive fraction of molecules taken up.